\documentclass[amsfonts,amsmath,prd,preprint,nofootinbib]{revtex4}%[aps,amsfonts,amsmath,prd,preprint,nofootinbib]{revtex4}

\newcommand{\beq}{\begin{equation}}
\newcommand{\eeq}{\end{equation}}

\usepackage{epsfig,bbm,cancel,ulem}
\usepackage{xcolor}
\usepackage[breaklinks=true]{hyperref}
\usepackage{graphicx}
\usepackage{amsmath}
\usepackage{amssymb}

\usepackage{wrapfig}
\usepackage{multirow}%
\usepackage{wasysym}
\usepackage[mathscr]{eucal}
\usepackage{amsfonts}%
\usepackage{amsthm}%
\usepackage{mathrsfs}%
\usepackage[title]{appendix}%
\usepackage{xcolor}%
\usepackage{textcomp}%
\usepackage{booktabs}%
\usepackage{algorithm}%
\usepackage{algorithmicx}%
\usepackage{algpseudocode}%
\usepackage{listings}%
\usepackage{subcaption}
\usepackage{tikzpagenodes}
\usepackage{tikz}
\usetikzlibrary{calc}
\usepackage{mwe}
\usepackage{setspace}
\usepackage{enumitem}
\usepackage{tabularray}
\usepackage{tablefootnote}
\usepackage{float}
\usepackage{xcolor}
\begin{document}

\title{Gravitational lensing by non-self-intersecting vortons}%{Weak-field Gravity and Lensing of Circular Chiral Vorton}

\author{Leonardus B.~Putra}
\email{leonardus.putra@exeter.ox.ac.uk, leonardus.brahmantyo@ui.ac.id}
\affiliation{Mathematical Institute, University of Oxford, Radcliffe Observatory, Andrew Wiles Building, Woodstock Rd, Oxford OX2 6GG}
\author{H.~S.~Ramadhan\footnote{Corresponding author.}}
\email{hramad@sci.ui.ac.id}

\affiliation{Departemen Fisika, FMIPA, Universitas Indonesia, Depok, 16424, Indonesia.}

\def\changenote#1{\footnote{\bf #1}}

\begin{abstract}
We investigate the gravitational lensing signatures of vorton configurations, considering the circular vorton, the Kibble–Turok vorton, and a newly proposed class that incorporates simultaneous excitations of the first, second, and third harmonic modes. Working within the weak-field and thin-lens approximations, we demonstrate that circular vortons produce a sharp lensing discontinuity that separates two regions with qualitatively distinct distortions. The corresponding Einstein ring co-exists alongside an almost undistorted source image. This effect is significantly amplified in the case of non-circular vortons, where asymmetries and higher-harmonic deformations amplify the discontinuity and lead to complex image structures. These distinctive lensing patterns offer potential discriminants between different vorton configurations, suggesting that future high-resolution surveys may provide a novel window into the microphysics of current-carrying cosmic strings. %—such as LSST, Euclid, and gravitational-wave observatories like LISA—may provide a novel window into the microphysics of current-carrying cosmic strings.
%We present exact solutions to the Nambu-Goto equations for thin vortons stabilized by chiral currents. The solutions describe a class of non-self-intersecting, stationary loops with arbitrary shapes. In addition to the trivial circular and the Kibble-Turok vortons, we also derive a two-parameter family that incorporates the first, second, and third harmonic modes. We found that, in general, the vorton's constraints allow for constructing families of solutions with arbitrary harmonic modes. We further investigate the gravitational lensing effects associated with these solutions under the weak-field and thin-lens approximations. For circular vortons, the lensing exhibits a sharp discontinuity separating two regions with distinctly different distortions. The corresponding Einstein ring co-exist alongside an almost undistorted source image. This effect is significantly amplified in the case of non-circular vortons, highlighting their potential observational signatures.
%Within the weak-field and thin-lens approximations, we analyze how these current-carrying cosmic string loops act as lenses and generate distinctive optical signatures.

\end{abstract}

\maketitle
\thispagestyle{empty}
%\section{Introduction}
\setcounter{page}{1}

\section{Introduction}
\label{sec1}

Cosmic strings are linear topological defects that may have been formed during symmetry-breaking phase transitions in the early Universe. First proposed in the context of grand unified theories and later embedded in superstring theory, these objects have been studied extensively for their cosmological and astrophysical implications, ranging from gravitational wave production to distinctive lensing phenomena \cite{Kibble:1976sj, Vilenkin:1984ib, Hindmarsh:1994re, Copeland:2009ga, CSBookVilenkin:2000jqa}. Of particular interest are superconducting cosmic strings, introduced by Witten in 1985~\cite{Witten:1984eb}, which admit persistent currents that can stabilize closed loops into long-lived solitonic states known as {\it vortons}. Early analyses, for example by Davis and Shellard~\cite{Davis:1988jp, Davis:1988jq, DAVISShellardVorton1989209}, Peter~\cite{Peter:1992dw}, and Carter~\cite{Carter:1989xk}, established the theoretical viability of vorton configurations, though their cosmological abundance and stability remain topics of investigation.

Vortons are stabilized by the balance between string tension, centrifugal forces, and current-induced stresses, resulting in nontrivial loop geometries.  Their formation, dynamics, stability, and cosmological significance have extensively been studied, for example, in~\cite{Blanco-Pillado_PhysRevD.63.103513, Martins:1998gb, Battye:2008mm, Carter:1990sm, Lemperiere:2003yt, Carter:1993wu}. Their existence carries potentially profound consequences for cosmology: an excessive vorton population could overclose the Universe, impose constraints on high-energy theories, or provide novel observational signatures \cite{Brandenberger:1996zp, Martins:2008zz}. In parallel, numerical and analytical studies have been devoted to their construction and stability \cite{Lemperiere:2002en, Radu:2008pp}.

Unlike current-less string loops, whose spacetime metric is inherently non-stationary~\cite{Frolov:1989er} and require an external pressure to prevent oscillatory collapse~\cite{Hughes_PhysRevD.47.468, McManus_PhysRevD.47.1491}, vortons possess a stationary metric. In~\cite{Putra:2024zms} we examine the metric around a circular vorton stabilized by a chiral current in the weak-field limit. The resulting metric functions are analogous to the 4-potential generated by a circular current-carrying wire loop. Asymptotically, the metric reduces to that of a Kerr black hole with mass $M=4\pi R\mu$, with $R$ the radius and $\mu$ the tension. We found that for a typical GUT-scale string %, $R\sim10^5-10^6 R_{Sch}$ , 
the extremal Kerr bound is always saturated. This implies that, to a distant observer, a GUT-scale vorton would be indistinguishable from a Kerr naked singularity. 

The phenomenological significance of these results is reinforced by recent observational developments. The detection of a stochastic gravitational-wave background by pulsar timing array collaborations has revived cosmic strings as a compelling candidate source~\cite{NANOGrav:2023hfp, EPTA:2023xxk}. Constraints from ground-based detectors such as LIGO–Virgo–KAGRA~\cite{LIGOScientific:2021nrg}, together with forecasts for the space-based LISA mission~\cite{Auclair:2019wcv}, continue to narrow the window for the string tension $G\mu$. At the same time, forthcoming large-scale surveys (LSST, Euclid, and the SKA) are expected to provide adequate sensitivity to the subtle lensing signatures produced by cosmic string loops, and in particular by vortons~\cite{Chernoff:2007pd, Benabou:2023ghl}. These observational opportunities motivate the development of precise theoretical predictions for vorton lensing.

%Recently, there has been growing interest in studying photon orbits and the shadows of Kerr naked singularities~\cite{Charbulak:2018wzb, Nguyen:2023clb}. This surge of interest is partly driven by the hypothesis that violations of the extremal Kerr bound could serve as a potential signature of string theory~\cite{Gimon:2007ur}. In our previous work~\cite{Putra:2024zms}, we analyzed the gravitational lensing of circular vortons by solving the null geodesic equations. 
In this paper, we extend the gravitational lensing analysis of vortons by systematically comparing three classes of configurations: (i) the canonical circular vorton, (ii) Kibble–Turok–like asymmetric loops, and (iii) new hybrid configurations incorporating first, second, and third harmonic modes. This generalization, however, comes at the cost of losing the explicit form of the metric. To address this, we employ the thin-lens approximation. Using the weak-field and thin-lens approximations, we derive lensing equations and image maps, elucidating how discontinuities, Einstein rings, and asymmetric distortions scale with loop geometry and harmonic content. This paper is organized as follows. In Section~\ref{sec:vortonsolutions} we introduce the general solutions for vorton loops and discuss their non-self-intersection constraints. In Section~\ref{sec:lensing} we calculate the corresponding deflection vectors and magnifications of the thin lens formalism. We then illustrate the lensing images of Milky Way galaxy caused circular and non-circular vortons. Finally, in Section~\ref{sec:conc} we summarize our findings and provide concluding remarks.

\section{Vorton Solutions}
\label{sec:vortonsolutions}

In the thin-string approximation the dynamics of relativistic loops is governed by the Nambu-Goto equation~\cite{Nambu, Goto:1971ce}, whose general solutions can be expanded in the Fourier modes. By considering the first and third harmonics, Kibble and Turok~\cite{Kibble:1982cb, Turok:1984cn} found a class of two-parameter family of smooth exact closed-loop solutions that never self-intersect. Generalizations to a third-parameter family or involving arbitrary number of harmonics were discussed in~\cite{Chen:1987ve, DeLaney:1989je}.

\subsection{Chiral String Loops}

The position of a string over time moving through spacetime is described by the three-vector $\vec{x}(\zeta,\tau)$, parameterised by one spacelike coordinate $\eta$ and one timelike coordinate $\tau$. Reparameterisation invariance allow us to choose $\tau=t$. This choice is standard in analyses of string dynamics, and we shall adopt it here as well. In this gauge, the string position vector obeys the flat-spacetime Nambu-Goto equations in conformal gauge:
\begin{eqnarray}
\label{eq:StringEq}
    \ddot{\Vec{x}}-\vec{x}''&=&0,\nonumber\\
\dot{\Vec{x}}\cdot\vec{x}'&=&0,\nonumber\\
\dot{\Vec{x}}^2+\vec{x}'^2&=&1,
\end{eqnarray}
where the prime ($'$) and dot ($\Dot{}$) correspond to derivative to $\zeta$ and $t$, respectively. The general solution to these equations is given by
\begin{equation}
\label{NGSol}
    \vec{x}=\frac{1}{2}\left[\vec{a}(\zeta-t)+\vec{b}(\zeta+t)\right],
\end{equation}
where $\vec{a}$ and $\vec{b}$ are traveling wave functions satisfying $\vec{a}'^2=\vec{b}'^2=1$.

Vorton is a superconducting string loop whose current generates angular momentum that counteracts its tendency to collapse. The current can either be electromagnetic or chiral. Carter and Peter showed that in the chiral limit the superconducting string's dynamics can be simplified~\cite{Carter:1999hx}. They proposed the following action
\begin{equation}
    S=\int d^2\zeta \left(-\mu\sqrt{-\gamma}+\frac{1}{2}\sqrt{-\gamma}\gamma^{ab}\varphi_{,a}\varphi_{,b}\right),
\end{equation}
where $\varphi$ is the internal phase field from Witten's superconducting string model. Here, $\gamma_{ab}$ is the induced metric of the worldsheet submanifold,
\begin{eqnarray}
    \gamma_{ab}=g_{\mu\nu}\frac{dx^\mu}{dy^a}\frac{dx^\nu}{dy^b},
\end{eqnarray}
with $a,b,...=1,2$ and $\left(y^1,y^2\right)=\left(t,\zeta\right)$. The determinant is $\gamma$. The chiral condition $\varphi_{,a}\varphi^{,a}=0$ acts as a constraint, and the equations of motion (EoM) are
\begin{equation}
\partial_a\left(\mathcal{T}^{ab}x^\nu_{,b}\right)=0,\ \ \ \ \ \ 
\partial_a\left(\sqrt{-\gamma}\gamma^{ab}\varphi_{,b}\right)=0,
\end{equation}
where
\begin{equation}
\mathcal{T}^{ab}\equiv\sqrt{-\gamma}\left(\mu \gamma^{ab}+\theta^{ab}\right),\ \ \ \ \theta_{ab}\equiv\varphi_{,a}\varphi_{,b},
\end{equation}
are the string worldsheet and
charge carrier energy-momentum tensors, respectively. 

It is convenient to choose a gauge so that% the worldsheet energy-momentum tensor has the form
\begin{equation}
\mathcal{T}^{ab}=\mu\eta^{ab},
\end{equation}
which is satisfied for the chiral case. In this gauge, %the energy-momentum tensor is given by
%\begin{equation}
%\label{emtensorchiral}
%    T^{\mu\nu}=\mu\int d\eta \left(\Dot{x}^\mu\Dot{x}^\nu-x'^\mu x'^\nu\right)\delta^{(3)}\left(\Vec{y}-\Vec{x}(t,\eta)\right).
%\end{equation}
the EoM becomes similar to that of Nambu-Goto strings Eq.~\eqref{NGSol},
\begin{equation}
    \ddot{\vec{x}}-\vec{x}''=0,\ \ \ \ \ \ \ \ \ \ddot{\varphi}-\varphi''=0,
\end{equation}
 %and
 %\begin{equation}
 %    \Ddot{\varphi}-\varphi''=0,
 %\end{equation}
whose solutions are
\begin{eqnarray}
\label{ChiralSol}
    \vec{x}=\frac{1}{2}(\vec{a}(\zeta-t)+\vec{b}(\zeta+t)),\ \ \ \
%\end{equation}
%and
%\begin{equation}
\varphi(\zeta,t)=F(\zeta+t),
\end{eqnarray}
with $F$ some function of $\zeta+t$. For the chiral string case, however, the constraints become
\begin{equation}
\label{ChiralConsa}
|\vec{a}'|=1,\ \ \ \ \ 
%\end{equation}
%and
%\begin{equation}
%\label{ChiralConsb}
|\vec{b}'|^2\equiv k^2=1-\frac{4F'^2}{\mu}.
\end{equation}

The general expansion of the coefficients are
\begin{eqnarray}
    \vec{a}(\zeta_-)&=&2\sum_{n=1}^\infty\left(\vec{A}_n^-\sin{\left(\frac{2\pi n \zeta_-}{L}\right)}+\vec{B}_n^-\cos{\left(\frac{2\pi n \zeta_-}{L}\right)}\right),\nonumber\\
%\end{equation}
%and
%\begin{equation}
    \vec{b}(\zeta_+)&=&2\sum_{n=1}^\infty\left(\vec{A}_n^+\sin{\left(\frac{2\pi n \zeta_+}{L}\right)}+\vec{B}_n^+\cos{\left(\frac{2\pi n \zeta_+}{L}\right)}\right),
\end{eqnarray}
with $\zeta_\pm=\zeta\pm t$. Expanding the traveling waves in Fourier modes with $L\equiv2\pi R$ and %$\zeta_\pm\equiv\zeta\pm t$,
%\eqref{ChiralConsa} leads to the following conditions:
%\begin{equation}
%\begin{split}
%        \frac{16\pi^2}{L^2}\sum_{nm}nm\left[\vec{A}_n^-\cdot \vec{A}_m^-\cos{\left(\frac{2\pi n\zeta_-}{L}\right)}\cos{\left(\frac{2\pi m\zeta_-}{L}\right)}\right.\\
%        -2\vec{A}_n^-\cdot \vec{B}_m^-\cos{\left(\frac{2\pi n\zeta_-}{L}\right)}\sin{\left(\frac{2\pi m\zeta_-}{L}\right)}\\+\left.
%        \vec{B}_n^-\cdot \vec{B}_m^-\sin{\left(\frac{2\pi n\zeta_-}{L}\right)}\sin{\left(\frac{2\pi m\zeta_-}{L}\right)}\right]=1,
%\end{split}
%\end{equation}
%and
%\begin{equation}
%\begin{split}
%        \frac{16\pi^2}{L^2}\sum_{nm}nm\left[\vec{A}_n^+\cdot \vec{A}_m^-\cos{\left(\frac{2\pi n\zeta_+}{L}\right)}\cos{\left(\frac{2\pi m\zeta_+}{L}\right)}\right.\\
%        -2\vec{A}_n^+\cdot \vec{B}_m^+\cos{\left(\frac{2\pi n\zeta_+}{L}\right)}\sin{\left(\frac{2\pi m\zeta_+}{L}\right)}\\+\left.
%        \vec{B}_n^+\cdot \vec{B}_m^+\sin{\left(\frac{2\pi n\zeta_+}{L}\right)}\sin{\left(\frac{2\pi m\zeta_+}{L}\right)}\right]=k^2.
%\end{split}
%\end{equation}
integrating them over $0\leq\zeta_\pm\leq L$ by defining new coefficients  $\vec{A}_n^\pm\equiv\vec{a}_n^\pm L/4\pi$, $\vec{B}_n^\pm\equiv\vec{b}_n^\pm L/4\pi$ yield
%\begin{equation}
%    \sum_{n=1}^\infty n^2\left(|\Vec{A}_n^-|^2+|\Vec{B}_n^-|^2\right)=\frac{L^2}{8\pi^2},\ \ \ \ \ \  
%    \sum_{n=1}^\infty n^2\left(|\Vec{A}_n^+|^2+|\Vec{B}_n^+|^2\right)=\frac{L^2}{8\pi^2}k^2.
%\end{equation}
\begin{equation}
\label{reduceda}
    \sum_{n=1}^\infty n^2\left(|\Vec{a}_n^-|^2+|\Vec{b}_n^-|^2\right)=2,\ \ \ \ 
%\label{reducedb}
    \sum_{n=1}^\infty n^2\left(|\Vec{a}_n^+|^2+|\Vec{b}_n^+|^2\right)=2k^2.
\end{equation}
For the Nambu-Goto strings, we have $k=1$, while for the vorton solutions, we have $k=0$.

\subsection{Circular Vorton}

It is easy to show that the following parametrizations 
\begin{eqnarray}
\vec{a}_n^-&=&\frac{\delta_{nm}}{m}\left(0,\cos{\phi},\sin{\phi}\right),\ \ \ \ \ \ 
%\end{equation}
%\begin{equation}
\vec{b}_n^-=\frac{\delta_{nm}}{m}\left(1,0,0\right),\\
%\end{equation}
%and
%\begin{equation}
\vec{a}_n^+&=&k\frac{\delta_{nm}}{m}\left(0,\cos{\phi},\sin{\phi}\right),\ \ \ \
%\end{equation}
%\begin{equation}
\vec{b}_n^+=k\frac{\delta_{nm}}{m}\left(1,0,0\right),
\end{eqnarray}
with $m$ integers are solutions of Eqs.~\eqref{reduceda}.
Following the convention used in Ref.~\cite{Davis_PhysRevD.62.083516}, and rescaling $R\rightarrow mR$, we write
\begin{eqnarray}
\label{acircloop}
\vec{a}(q)&=&R\left(\cos{\frac{q}{R}},\cos{\phi}\sin{\frac{q}{R}},\sin{\phi}\sin{\frac{q}{R}}\right),\nonumber\\
\vec{b}(\eta)&=&k R\left(\cos{\frac{\eta}{R}},\cos{\phi}\sin{\frac{\eta}{R}},\sin{\phi}\sin{\frac{\eta}{R}}\right),
%\vec{a}(q)&=&\frac{R}{m}\left(\cos{\frac{mq}{R}},\cos{\phi}\sin{\frac{mq}{R}},\sin{\phi}\sin{\frac{mq}{R}}\right),\nonumber\\
%\end{equation}
%\begin{equation}
%\label{bcircloop}
%\vec{b}(\eta)&=&k \frac{R}{m}\left(\cos{\frac{m\eta}{R}},\cos{\phi}\sin{\frac{m\eta}{R}},\sin{\phi}\sin{\frac{m\eta}{R}}\right),
\end{eqnarray}
with $q\equiv\zeta_+= t+\zeta$, $\eta\equiv\zeta_-= t-\zeta$, and $\zeta$ the string's worldsheet spacelike coordinate. %We could rescale $R$ as $R\rightarrow mR$, so that the choice of $m=1$ is usfficient to describe all solutions in the family. 

This family of solutions describes a circular string loop oscillating between radius of $(1-k)R/2$ and $(1+k)R/2$, tilted with an angle $\phi$ around the $x$ axis. The Nambu-Goto limit of this family can be found by setting $k=1$, 
\begin{equation}
\label{circNGsol}
    \vec{r}=R\cos{\frac{t}{R}}\left(\cos{\frac{\zeta}{R}},\cos{\phi}\sin{\frac{\zeta}{R}},\sin{\phi}\sin{\frac{\zeta}{R}}\right),
\end{equation}
describing an oscillating circular string loop of maximum radius $R$ without current. The stationary, or ``vorton'', limit arises by setting $k=0$ and rescaling $R\rightarrow 2R$, leading to
\begin{equation}
\label{circvortonsol}
\vec{r}=R\left(\cos{\frac{q}{2R}},\cos{\phi}\sin{\frac{q}{2R}},\sin{\phi}\sin{\frac{q}{2R}}\right),
\end{equation}
which describes a circular vorton with radius $R$.

\subsection{Kibble-Turok Vorton}

Kibble and Turok\cite{Kibble:1982cb, Turok:1984cn} presented a class of string loop solutions by considering the first and third harmonics. If the string is chiral, then we set $k=1$ in \eqref{ChiralConsa}. This Kibble-Turok chiral loop reads
\begin{eqnarray}
\vec{a}_1^-&=&\left(1-\kappa,0,0\right),\ \ \ \
\vec{b}_1^-=\left(0,-(1-\kappa),-2\sqrt{\kappa(1-\kappa)}\right),\nonumber\\
\vec{a}_3^-&=&\left(\frac{1}{3}\kappa,0,0\right),\ \ \ \ \ 
\vec{b}_3^-=\left(0,-\frac{1}{3}\kappa,0\right),\nonumber\\\vec{a}_1^+&=&\left(1,0,0\right),\ \ \ \ \ \ \ \ \
\vec{b}_1^+=\left(0,-\cos{\phi},-\sin{\phi}\right),
\end{eqnarray}
where $0<\kappa\leq 1$ and $-\pi \leq \phi \leq \pi$ are constant parameters. We can generalize this result for the case of $k\neq 1$ by setting
\begin{equation}
\vec{a}_1^+=k\left(1,0,0\right),\ \ \ 
\vec{b}_1^+=k\left(0,-\cos{\phi},-\sin{\phi}\right).
\end{equation}
Using the same convention as in Eq.~\eqref{acircloop} and setting $L=2\pi R$, we get
\begin{eqnarray}
\vec{r}&=&\frac{R}{2}\bigg[(1-\kappa)\sin{\frac{q}{R}}+\frac{1}{3}\kappa \sin{\frac{3q}{R}}+k\sin{\frac{\eta}{R}},\nonumber\\
&&-\left((1-\kappa)\cos{\frac{q}{R}}+\frac{1}{3}\kappa\cos{\frac{3q}{R}}+k\cos{\phi}\cos{\frac{\eta}{R}}\right),\nonumber\\
&&-\left(2\sqrt{\kappa(1-\kappa)}\cos{\frac{q}{R}}+k\sin{\phi}\cos{\frac{\eta}{R}}\right)\bigg].    
\end{eqnarray}
It should be noted that the limit of $\kappa=0$ and $\phi=0$ brings the solution back to the circular case.
\begin{figure}
     \centering
     \begin{subfigure}[b]{0.24\textwidth}
         \centering
\includegraphics[width=\textwidth]{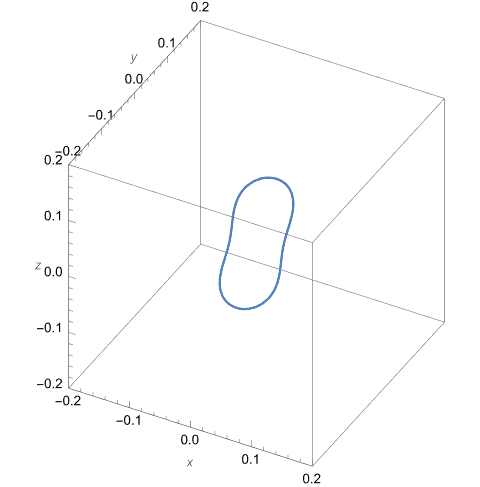}
    \caption{$\kappa=0.35$}
         \label{fig:3dk035}
     \end{subfigure}
     \hfill
     \begin{subfigure}[b]{0.24\textwidth}
         \centering
\includegraphics[width=\textwidth]{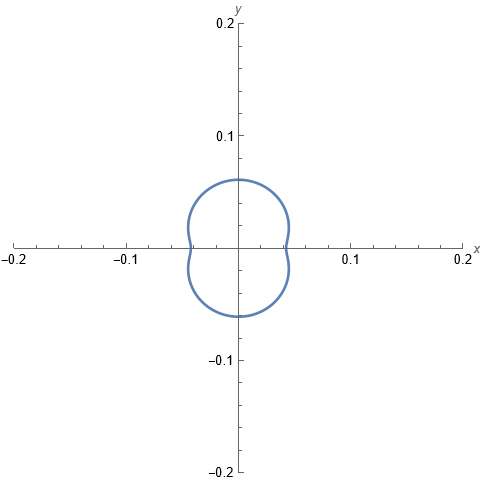}
    \caption{$\kappa=0.35$}
         \label{fig:2dk035}
     \end{subfigure}
     \hfill
     \begin{subfigure}[b]{0.24\textwidth}
         \centering
\includegraphics[width=\textwidth]{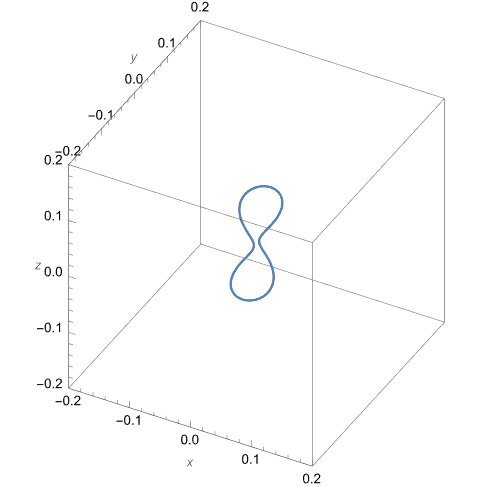}
        \caption{$\kappa=0.7$}
         \label{fig:3dk070}
     \end{subfigure}
     \hfill
     \begin{subfigure}[b]{0.24\textwidth}
         \centering
\includegraphics[width=\textwidth]{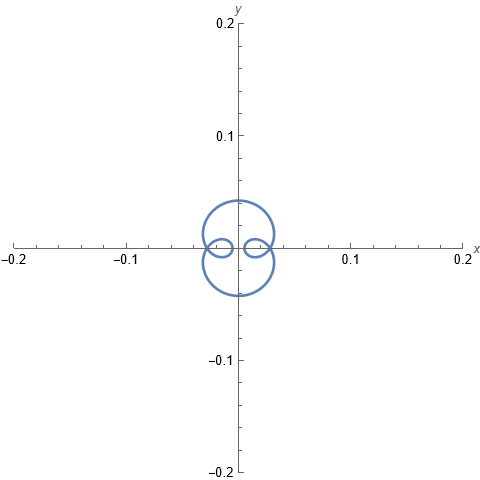}
\caption{$\kappa=0.7$}
         \label{fig:2dk070}
     \end{subfigure}
        \caption{Parametric 3D (a, c) and top-view (b, d) curves of Kibble-Turok vorton solutions of various $\kappa$.}
    \label{fig:turokvortonsol}
\end{figure}

 The Kibble-Turok vorton can be obtained by taking $k=0$ and $R\rightarrow 2R$, which yields
\begin{equation}
\label{TurokVorton}
%\begin{split}
\vec{r}=R\left[(1-\kappa)\sin{\frac{q}{2R}}+\frac{1}{3}\kappa \sin{\frac{3q}{2R}},-(1-\kappa)\cos{\frac{q}{2R}}-\frac{1}{3}\kappa\cos{\frac{3q}{2R}},-2\sqrt{\kappa(1-\kappa)}\cos{\frac{q}{2R}}\right].    
%\end{split}
\end{equation}
Since the dependence on $\phi$ couples to $k$, the Kibble-Turok vorton is parametrized by only one paramater, $\kappa$. In Fig.~\ref{fig:turokvortonsol} we show Kibble-Turok vorton profiles for several $\kappa$.

\subsection{The 123-harmonic Vorton}

%While the Kibble-Turok solution assumes only the first and third harmonics to be non-zero, we can include the second harmonics as well to create a solution where all the first three harmonics are present. Here, we choose

While the original Kibble-Turok solution considers only the first and third non-zero harmonics, our vorton loop formalism allows the inclusion of all harmonic modes. Here, we construct a vorton loop that includes the first, second, and third harmonics, referred to as {\it the $123$-vorton} for short, given by

\begin{eqnarray}
\label{123}
\vec{a}_1^-&=&\left(\sqrt{\beta}(1-\kappa),0,0\right),\ \ \ \ \ \ \ \ \ \ \ \ \
%\end{equation}
%\begin{equation}
\vec{b}_1^-=\left(0,-\sqrt{\beta}(1-\kappa),-2\sqrt{\beta\kappa(1-\kappa)}\right),\nonumber\\
\vec{a}_2^-&=&\left(0,\frac{1}{\sqrt{2}}\sqrt{1-2\beta-\kappa^2},0\right),\ \ \ \  
\vec{b}_2^-=\left(0,\sqrt{\frac{\beta}{2}},\kappa \sqrt{\frac{\beta}{2}}\right),\nonumber\\
\vec{a}_3^-&=&\left(\frac{1}{3}\kappa,0,0\right),\ \ \ \ \ \ \ \ \ \ \ \ \ \ \ \ \ \ \ \ \ \ \
\vec{b}_3^-=\left(0,-\frac{1}{3}\kappa,0\right),
\end{eqnarray}
with the $\zeta_+$ terms remaining as in the Kibble-Turok solution
\begin{equation}
\vec{a}_1^+=k\left(1,0,0\right),\ \ \ \ \ \ \ \ \ \ \
\vec{b}_1^+=k\left(0,-\cos{\phi},-\sin{\phi}\right).
\end{equation}
As in the Kibble-Turok vorton, we have $0<\kappa\leq 1$ and $-\pi\leq\phi\leq\pi$. The parameter $\beta$ is introduced to incorporate the second harmonic into the solution, with $0<\beta\leq\frac{1}{2}(1-\kappa^2)$. This solution, however, is not an exact generalization of the Kibble-Turok solution, as the Kibble-Turok solution cannot be fully recovered by setting a specific value of $\beta$. 
\begin{figure}
     \centering
     \begin{subfigure}[b]{0.24\textwidth}
         \centering
\includegraphics[width=\textwidth]{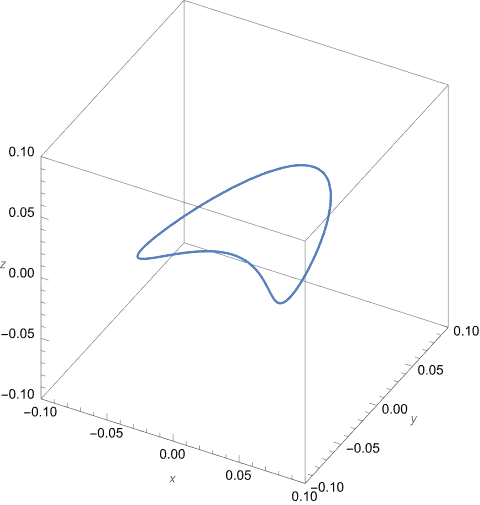}
         \caption{$\kappa=0$, $\beta=0.5$}
         \label{fig:3dk0b05}
     \end{subfigure}
     \hfill
     \begin{subfigure}[b]{0.24\textwidth}
         \centering
\includegraphics[width=\textwidth]{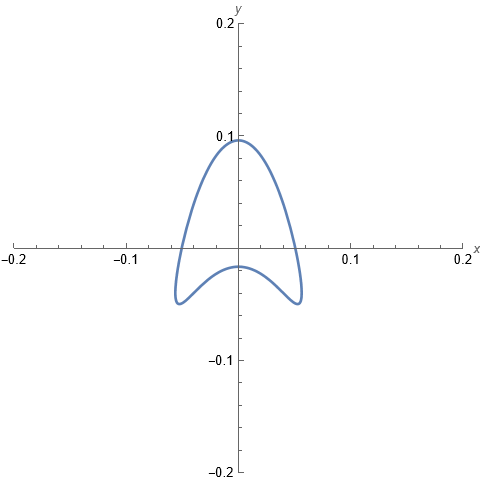}
         \caption{$\kappa=0$, $\beta=0.5$}
         \label{fig:2dk0b05}
     \end{subfigure}
     \hfill
     \begin{subfigure}[b]{0.24\textwidth}
         \centering
\includegraphics[width=\textwidth]{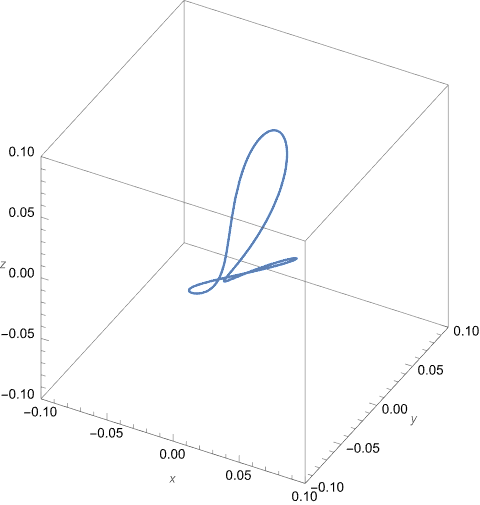}
\caption{$\kappa=0.5$, $\beta=0.2$}
         \label{fig:3dk05b02}
     \end{subfigure}
     \hfill
     \begin{subfigure}[b]{0.24\textwidth}
         \centering
\includegraphics[width=\textwidth]{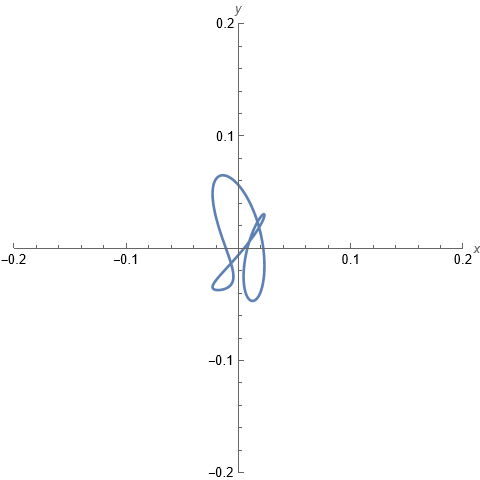}
    \caption{$\kappa=0.5$, $\beta=0.2$}
         \label{fig:2dk05b02}
     \end{subfigure}
        \caption{Parametric 3D (a, c) and top-view (b, d) curves of the $123$ vorton solutions of various $\kappa$ and $\beta$.}
\label{fig:123vortonsol}
\end{figure}

The $123$-vorton then describes a class of two-parameter family of solution given by
\begin{eqnarray}
\label{123Vorton}
\vec{r}&=&R\bigg[\sqrt{\beta}(1-\kappa)\sin{\frac{q}{2R}}+\frac{1}{3}\kappa \sin{\frac{3q}{2R}},\nonumber\\
    &&-\left(\sqrt{\beta}(1-\kappa)\cos{\frac{q}{2R}}-\frac{1}{\sqrt{2}}\sqrt{1-2\beta-\kappa^2}\sin{\frac{q}{R}}-\sqrt{\frac{\beta}{2}}\cos{\frac{q}{R}}+\frac{1}{3}\kappa\cos{\frac{3q}{2R}}\right),\nonumber\\
&&\kappa\sqrt{\frac{\beta}{2}}\cos{\frac{q}{R}}-2\sqrt{\kappa(1-\kappa)}\cos{\frac{q}{2R}}\bigg],    
\end{eqnarray}
parametrized by $\kappa$ and $\beta$. The solutions are shown in Fig.~\ref{fig:123vortonsol} for various values of $\kappa$ and $\beta$.

\subsection{Self-intersection and Additional Constraints}

Strings can intersect and interact with themselves. Imposing a non-self-intersection condition introduces additional constraints on our solutions. %, which requires the field coupling treatment of the problem. 
% This interaction is not considered in the thin string approximation used here, and therefore the self intersection would be ecluded in the preceeding section. This requirement creates some additional constraints to the solution. 
The solution $\vec{r}$ will self-intersect if and only if $\exists \alpha\in \left(0,2\pi\right)\And \theta\in [0,2\pi]$ such that
\begin{equation}
\vec{r}(\theta+\alpha)=\vec{r}(\theta).
\end{equation}
The circular Nambu-Goto string loop and vorton solutions, Eq.~\eqref{circNGsol}-\eqref{circvortonsol}, are inherently circular and, therefore, do not self-intersect. It is worth noting that the definition of self-intersection used in this study is slightly different from that in~\cite{Davis_PhysRevD.62.083516}. Here, the self-intersection refers to configurations where the string continuously intersects itself at  fixed point (stationary intersection). In contrast, Ref.~\cite{Davis_PhysRevD.62.083516} defines the self-intersection to be configurations where the loop intersects itself at some specific time $t$ (dynamic intersection).

To ensure a non-self-intersecting Kibble-Turok vorton, we impose the additional constraint
\begin{equation}
 0<\kappa<3/4,   
\end{equation}
where $\kappa=0$ corresponds to the circular loop.

And for the $123$-vorton:
\begin{equation}
    \frac{3}{4\kappa}(\sqrt{\beta}-\kappa\sqrt{\beta}+\kappa)\neq1,\ \ \ \ \ \ \  
%\end{equation}
%and
%\begin{equation}
    1-2\beta-\kappa^2\neq 0.
\end{equation}

\section{Lensing of Vorton}
\label{sec:lensing}

\subsection{Thin Lens Formalism}

Gravitational lensing by string loop can be computed using the thin lens approximation, where lensing takes place at a specific moment in time $t_0$, alongside the weak-field approximation of linearized gravity, as utilized by De Laix and Vachaspati in~\cite{deLaix:1996vc}. This formalism can be split into two main components. The first is the deflection vector, which constructs the lensing image. The second component is the magnification, which defines the curve of infinite magnification (the critical curve) and quantifies the image scale.  

%\subsubsection{Deflection Vector}

\begin{figure}
	\centering
\includegraphics[width=0.75\textwidth]
	{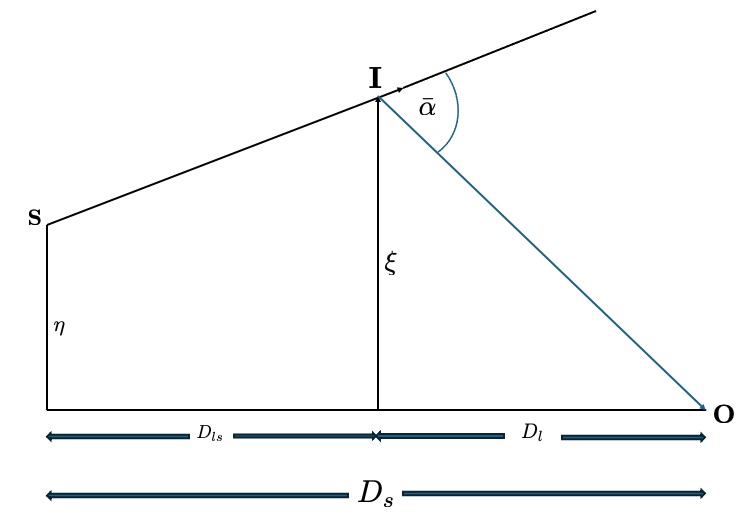}
	\caption[Lensing Diagram in Thin Lens Approximation]{Lensing Diagram in the Thin Lens Approximation~\cite{deLaix:1996vc}.}
	\label{fig:TLDiag}
\end{figure}

The deflection of photon from its flat spacetime trajectory is described by $\vec{\alpha}$, which represents the change in the photon’s velocity vector due to lensing. This can be expressed as
\begin{equation}
\Bar{\vec{\alpha}}=-4G\mu \int d\zeta \left(\frac{F_{\mu\nu}(\zeta,t)\gamma^\mu\gamma^\nu}{1-\dot{f_\parallel}}\frac{\Vec{f}_\perp}{f_\perp^2}\right)_{t=t_0},
\end{equation}
with
\begin{equation}
\label{Fmunulensing}
F_{\mu\nu}\equiv\dot{f}_\mu\dot{f}_\nu-f_\mu'f_\nu'-\eta_{\mu\nu}\dot{f}^2,
\end{equation}
$\gamma^\mu$ the four-velocity of the light ray, $f_\mu$ the parameterized coordinate of the string loop, and $t_0$ the solution of $f_\parallel(t_0,\zeta)=t_0$. The lensing is governed by the lens equation~\cite{Schneider:1992bmb}
\begin{equation}
    \vec{\eta}=\frac{D_s}{D_l}\vec{\xi}-D_{ls}\bar{\vec{\alpha}}(\vec{\xi}),
\end{equation}
where $D_l$ is the distance from observer to the lens, $D_s$ is the distance from the observer to source, and $D_{ls}$ the distance from the lens to source. The lens diagram is shown in Fig.\ref{fig:TLDiag}. Here we can rescale the angular deflection vector to better fit our defined parameters ($f_\mu$ and $\gamma_\mu$). We define
\begin{equation}
    \Vec{\alpha}(\Vec{x})=\Bar{\vec{\alpha}}(\Vec{\xi})\frac{D_{ls}D_l}{D_sR}.
\end{equation}
From Eq.~\eqref{Fmunulensing}, we have
\begin{equation}
%    \begin{split}
F_{\mu\nu}\gamma^\mu\gamma^\nu%&=\left(\Dot{f}_\mu\gamma^\mu\right)^2-\left(f_\mu'\gamma^\mu\right)^2-\gamma_\mu\gamma^\mu\Dot{f}^2\\
=\left(1-\Dot{\Vec{f}}\cdot\hat{\gamma}\right)^2-\left(-\Vec{f}'\cdot\hat{\gamma}\right)^2,
%    \end{split}
\end{equation}
with $\hat{\gamma}$ the spatial unit vector of $\gamma^{\mu}$. The vector $\vec{f}$ represents the displacement from the light ray’s position in the lensing plane (at $t_0$) to the parametric four-coordinate of the string loop. Taking the string loop's center of mass as the origin, the 3-vector $\vec{f}$ can be decomposed as 
$\vec{f}=\Vec{r}-\Vec{x}_0$,
where $\vec{r}$ is the coordinate of the string loop, and $\vec{x}_0$ is the light ray coordinate at $t_0$ (when the light ray hits the lensing plane). This vector $\vec{f}$ can be further decomposed (wrt to the optical axis) into the perpendicular and parallel components
\begin{equation}
\vec{f}_\perp=\vec{r}_\perp-\vec{\xi},\ \ \ \ \ \ \ \vec{f}_\parallel=\vec{r}\cdot\hat{\gamma}-\Vec{x}_0\cdot\hat{\gamma},
\end{equation}
with $\vec{\xi}$ the perpendicular component of $\vec{x}_0$. Thus, we have
\begin{equation}
\dot{f}_\parallel=\frac{d}{dt}\left(\vec{r}\cdot\hat{\gamma}\right)=\dot{r}_\gamma.
\end{equation}
Here $r_\gamma$ is the string loop vector component parallel to the optical axis. We then have
\begin{equation}
\label{vecalpha}
\vec{\alpha}=-4G\mu\frac{D_{ls}D_l}{RD_s}\int d\zeta \left[\frac{\left(1-\dot{r}_\gamma\right)^2-r_\gamma'^2}{1-\dot{r}_\gamma}\frac{\vec{r}_\perp-\vec{\xi}}{\left(\Vec{r}_\perp-\vec{\xi}\right)^2}\right]_{t=t_0}.
\end{equation}
We can re-parametrize $\vec{\xi}$ and $\vec{r}_\perp$ as
\begin{equation}
\vec{\xi}=R\left(x_1\hat{e}_1+x_2\hat{e}_2\right), \ \ \ \ \ \ \vec{r}_\perp=R\left(\frac{r_1}{R}\hat{e}_1+\frac{r_2}{R}\hat{e}_2\right).
\end{equation}
Choosing the optical axis to be $\hat{e}_3$ along with the light ray direction $\hat{\gamma}=\hat{e}_3$ (assuming small deflection angle), %and that the light ray direction parallel to it, $\hat{\gamma}=\hat{e}_3$, since we also assume the deflection angle to be small,
%\begin{equation}
%    \hat{\gamma}=\hat{e}_3.
%\end{equation}
we have
\begin{equation}
    \vec{\alpha}=C\vec{F},
\end{equation}
with
\begin{equation}
C\equiv8\pi G\mu \frac{D_{ls}D_l}{RD_s},
\end{equation}
a constant, and
\begin{equation}
\vec{F}\equiv F_1(x_1,x_2)\hat{e}_1+F_2(x_1,x_2)\hat{e}_2,
\end{equation}
the deflection vector, %is defined as the deflection vector, with components
where
\begin{eqnarray}
    F_1(x_1,x_2)&\equiv&-\frac{1}{2\pi}\int \frac{d\zeta}{R} \left[\frac{\left(1-\Dot{r}_\gamma\right)^2-r_\gamma'^2}{1-\Dot{r}_\gamma}\frac{\frac{r_1}{R}-x_1}{\left(\frac{r_1}{R}-x_1\right)^2+\left(\frac{r_2}{R}-x_2\right)^2}\right]_{t=t_0},\nonumber\\
%\end{equ}
%and
%\begin{equation}
    F_2(x_1,x_2)&\equiv&-\frac{1}{2\pi}\int \frac{d\zeta}{R} \left[\frac{\left(1-\Dot{r}_\gamma\right)^2-r_\gamma'^2}{1-\Dot{r}_\gamma}\frac{\frac{r_1}{R}-x_1}{\left(\frac{r_1}{R}-x_1\right)^2+\left(\frac{r_2}{R}-x_2\right)^2}\right]_{t=t_0}.
\end{eqnarray}
The lens equation therefore becomes
\begin{eqnarray}
\label{rescaledlenseq}
%    \begin{split}
        y_1&=&x_1-CF_1(x_1,x_2),\nonumber\\
        y_2&=&x_2-CF_2(x_1,x_2).
%    \end{split}
\end{eqnarray}
The rescaled coordinates $\vec{y}$ and $\vec{x}$ are the source and the image positions, respectively.

%\subsubsection{Magnification and Caustics}
% and $\vec{x}$ is the image position.

%The image mapping transformation \eqref{rescaledlenseq} has the Jacobian
%\begin{equation}
%    J_{ij}=\frac{\partial y_i}{\partial x_j}.
%\end{equation}
The magnification of the transformed (mapped) image is given by~\cite{Schneider:1992bmb}:
\begin{equation}
\label{magnificationdef}
%\begin{split}
    M=\frac{1}{\det{J}\left(x_1,x_2\right)},
%\end{split}
\end{equation}
where $J_{ij}\left(x_1,x_2\right)$ is the Jacobian of Eq.~\eqref{rescaledlenseq}.
Thus,
\begin{equation}
\label{JacobianDeterminantFinalForm}
    M=\frac{1}{\left(1-CF_{1,1}\right)\left(1-CF_{2,2}\right)-C^2F_{1,2}F_{2,1}},
\end{equation}
where $F_{i,j}\equiv\partial F_i/\partial x_j$. The critical curve, defined as the curve with infinite magnification, is  just the curve $(x_1,x_2)$ satisfying
\begin{equation}
\label{critcurveeq}
    \left(1-CF_{1,1}\right)\left(1-CF_{2,2}\right)-C^2F_{1,2}F_{2,1}=0.
\end{equation}
Since Eqs.~\eqref{rescaledlenseq} map $\vec{x}\rightarrow\vec{y}$, the caustic images are obtained by substituting the solution of Eq.~\eqref{critcurveeq} into Eqs.~\eqref{rescaledlenseq}.

\subsection{Circular Vorton and Nambu-Goto Loops}

\begin{figure}[H]
     \centering
     \begin{subfigure}[b]{0.3\textwidth}
         \centering
\includegraphics[width=\textwidth]{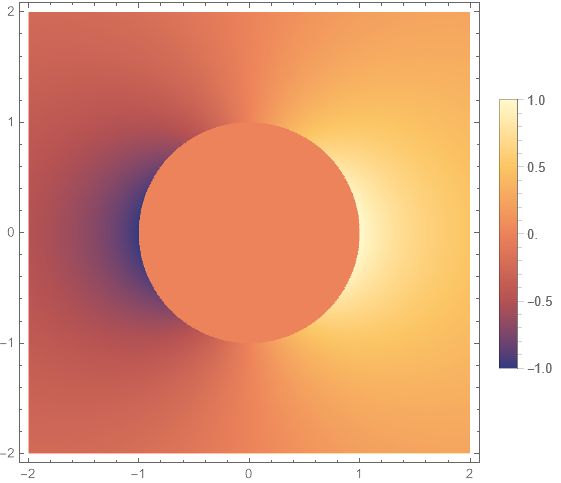}
         \caption{$\phi=0$}
         \label{fig:ccs0F1}
     \end{subfigure}
     \hfill
     \begin{subfigure}[b]{0.3\textwidth}

     \centering
\includegraphics[width=\textwidth]{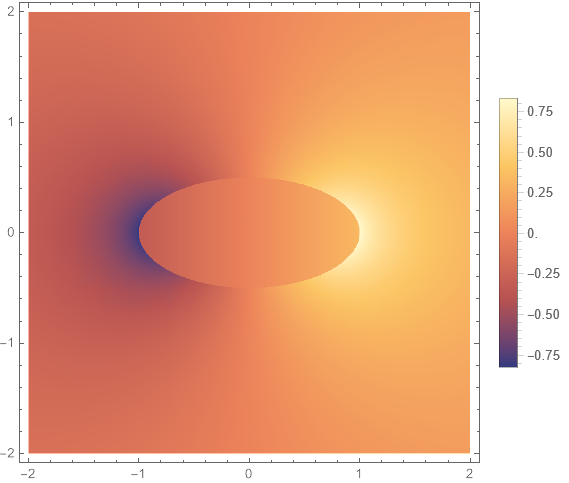}
         \caption{$\phi=\frac{\pi}{3}$}
         \label{fig:ccs60F1}
     \end{subfigure}
     \hfill
     \begin{subfigure}[b]{0.3\textwidth}
     
         \centering
\includegraphics[width=\textwidth]{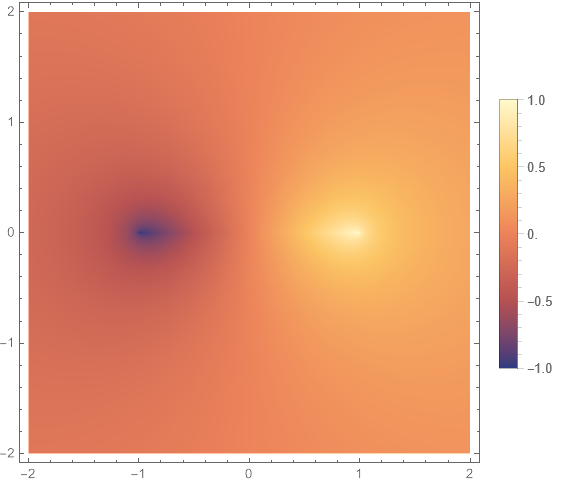}
         \caption{$\phi=\frac{\pi}{2}$}
         \label{fig:ccs90F1}
     \end{subfigure}
     \hfill
     \begin{subfigure}[b]{0.3\textwidth}
     
         \centering
\includegraphics[width=\textwidth]{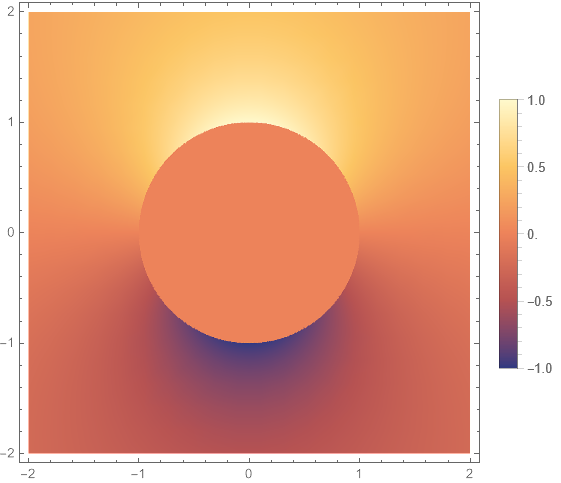}
         \caption{$\phi=0$}
         \label{fig:ccs0F2}
     \end{subfigure}
     \hfill
     \begin{subfigure}[b]{0.3\textwidth}
     
         \centering
\includegraphics[width=\textwidth]{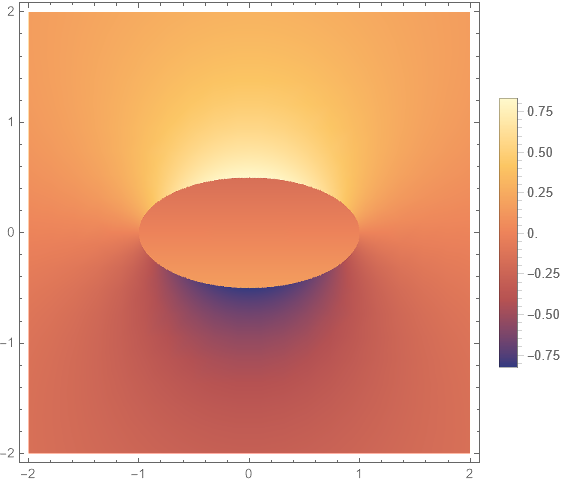}
         \caption{$\phi=\frac{\pi}{3}$}
         \label{fig:ccs60F2}
     \end{subfigure}
     \hfill
     \begin{subfigure}[b]{0.3\textwidth}

         \centering
\includegraphics[width=\textwidth]{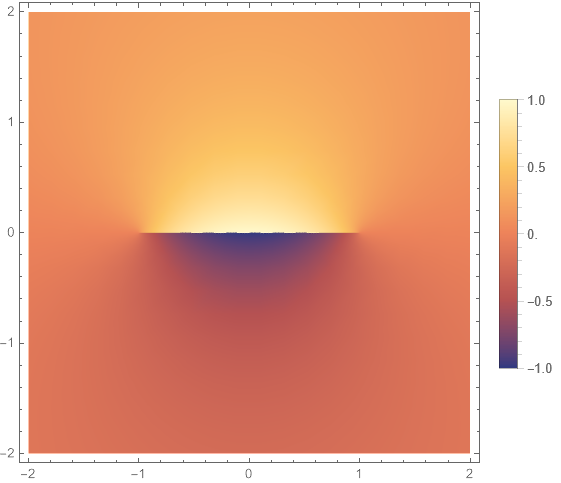}
         \caption{$\phi=\frac{\pi}{2}$}
         \label{fig:ccs90F2}
\end{subfigure}
\caption{Deflection vector component $F_1$ (a, b, c) and $F_2$ (d, e, f) from circular Nambu-Goto loop of various $\phi$, illustrated as a heatmap. The maximum and (magnitude of) minimum values of the $F_1$ and $F_2$ component are the same in the respective image. The brighter the color in the heatmap, the larger the deflection in the respective direction. For example, (a) shows large deflection to positive (negative) horizontal direction right to the left (right) of the string, and little to no deflection in the horizontal direction on top of the projection. The plot depicting $F_2$ shows the distribution of vertical deflection.}
\label{fig:F12ccs}
\end{figure}

\begin{figure}[H]
     \centering
     \begin{subfigure}[b]{0.24\textwidth}
         \centering
\includegraphics[width=\textwidth]{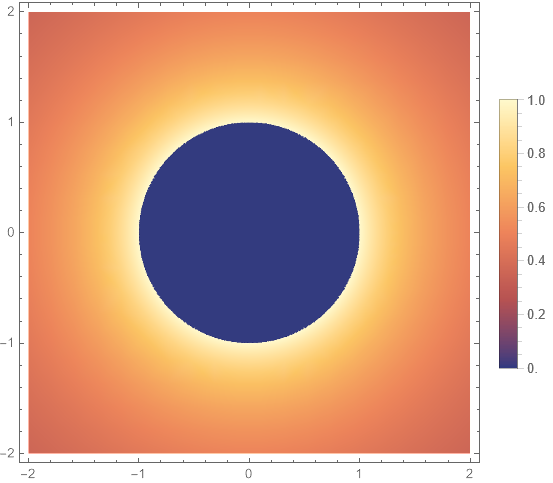}
         \caption{$\phi=0$}
         \label{fig:ccs0F}
     \end{subfigure}
     \hfill
     \begin{subfigure}[b]{0.24\textwidth}
         \centering
\includegraphics[width=\textwidth]{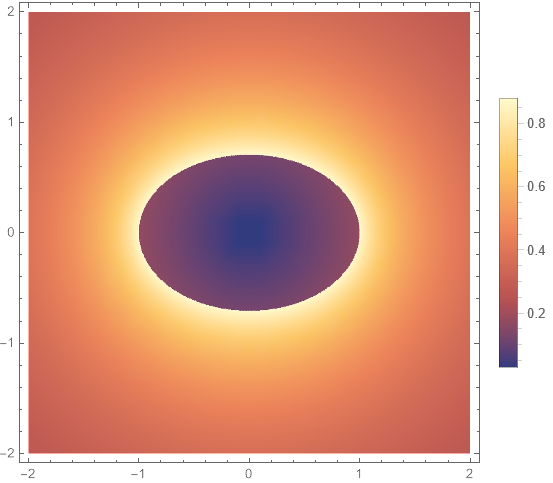}
         \caption{$\phi=\frac{\pi}{4}$}
         \label{fig:ccs45F}
     \end{subfigure}
     \hfill
     \begin{subfigure}[b]{0.24\textwidth}
         \centering
\includegraphics[width=\textwidth]{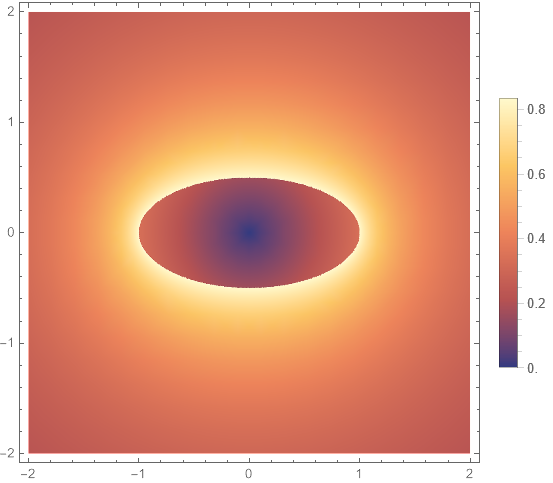}
         \caption{$\phi=\frac{\pi}{3}$}
         \label{fig:ccs60F}
     \end{subfigure}
     \hfill
     \begin{subfigure}[b]{0.24\textwidth}
         \centering
\includegraphics[width=\textwidth]{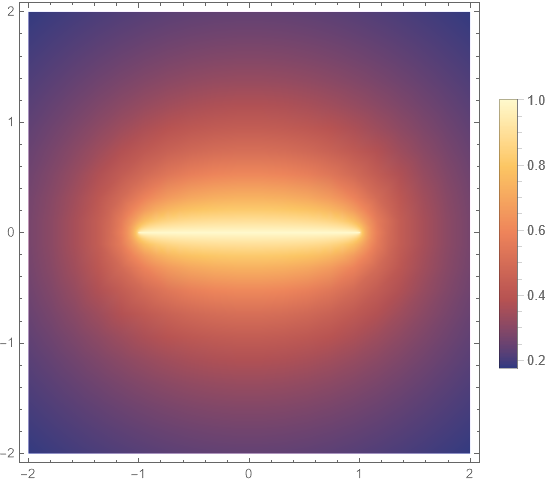}
         \caption{$\phi=\frac{\pi}{2}$}
         \label{fig:ccs90F}
     \end{subfigure}
        \caption{Deflection vector magnitude $|\Vec{F}|=\sqrt{F_1^2+F_2^2}$ from circular Nambu-Goto loop for several $\phi$. The brighter the color, the larger the angular deflection magnitude is. Fig.~(a) shows the deflection around the outer side of the ring projection is the largest, while no deflection occurs inside the string projection.}
        \label{fig:Fccs}
\end{figure}

\begin{figure}[H]
     \centering
     \begin{subfigure}[b]{0.3\textwidth}
         \centering
\includegraphics[width=\textwidth]{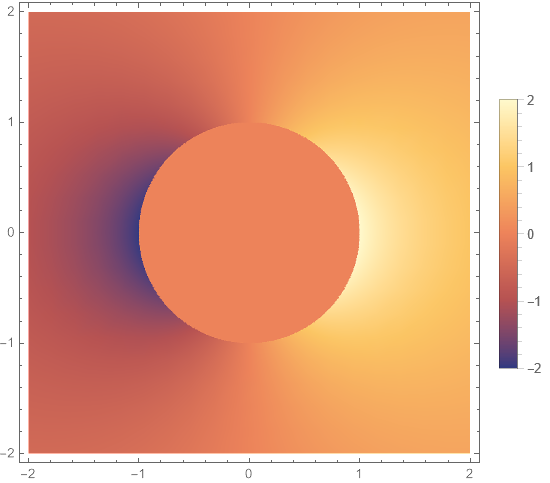}
         \caption{$\phi=0$}
         \label{fig:cvort0F1}
     \end{subfigure}
     \hfill
     \begin{subfigure}[b]{0.3\textwidth}
     
         \centering
\includegraphics[width=\textwidth]{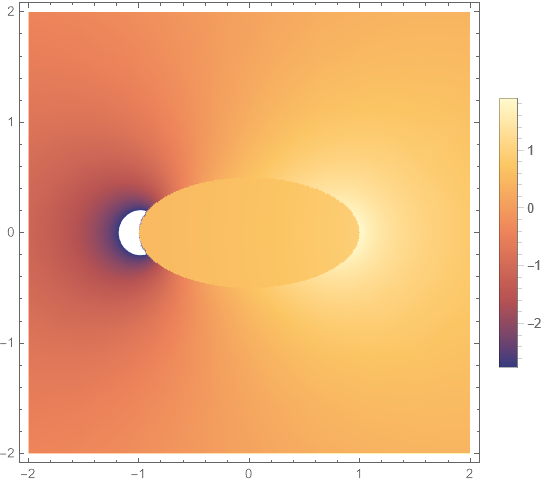}
         \caption{$\phi=\frac{\pi}{3}$}
         \label{fig:cvort60F1}
     \end{subfigure}
     \hfill
     \begin{subfigure}[b]{0.3\textwidth}
     
         \centering
\includegraphics[width=\textwidth]{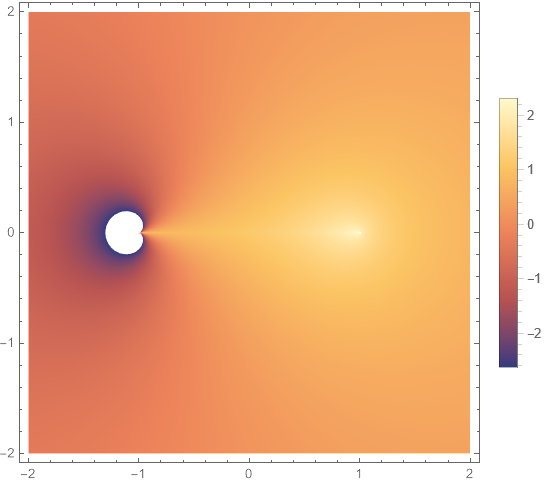}
         \caption{$\phi=\frac{\pi}{2}$}
         \label{fig:cvort90F1}
     \end{subfigure}
     \hfill
     \begin{subfigure}[b]{0.3\textwidth}
     
         \centering
\includegraphics[width=\textwidth]{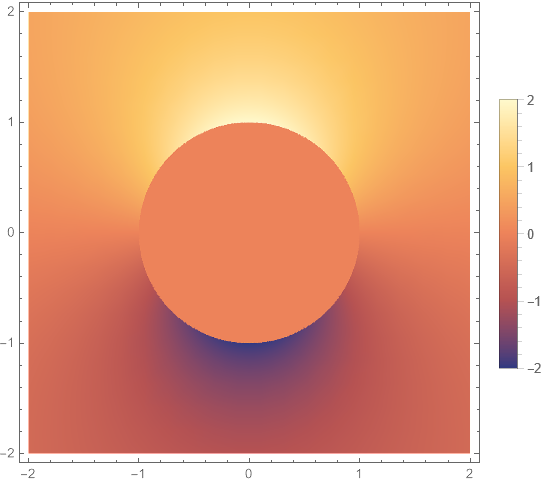}
         \caption{$\phi=0$}
         \label{fig:cvort0F2}
     \end{subfigure}
     \hfill
     \begin{subfigure}[b]{0.3\textwidth}

         \centering
\includegraphics[width=\textwidth]{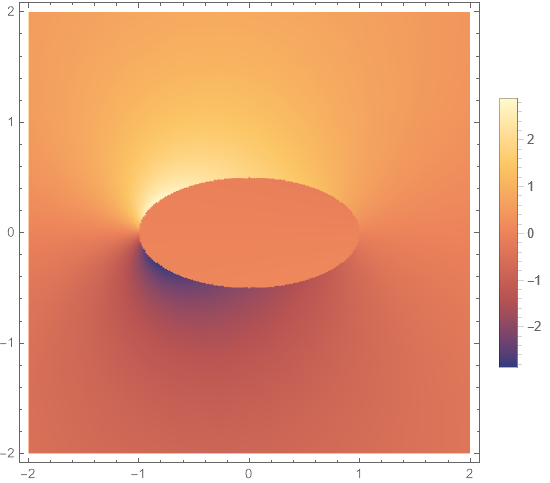}
         \caption{$\phi=\frac{\pi}{3}$}
         \label{fig:cvort60F2}
     \end{subfigure}
     \hfill
     \begin{subfigure}[b]{0.3\textwidth}

         \centering
\includegraphics[width=\textwidth]{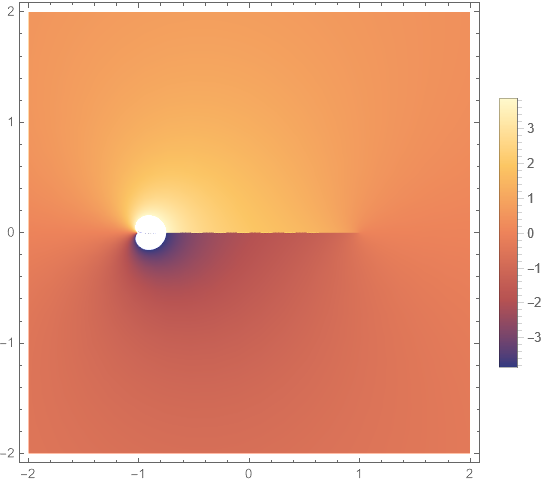}
         \caption{$\phi=\frac{\pi}{2}$}
         \label{fig:cvort90F2}
     \end{subfigure}
        \caption{Deflection vectors $F_1$ (a, b, c) and $F_2$ (d, e, f) from circular vorton for several $\phi$. The white spots surrounded by dark color as seen in (b, c) and the lower half of (f), should be interpreted as divergence in negative value. Figs.~(b, c) shows a large positive horizontal deflection right on the left side of the projection, and (f) shows large negative (positive) vertical deflection right on top (bottom) of the left end of string projection.}
        \label{fig:F12cvort}
\end{figure}

\begin{figure}[H]
     \centering
     \begin{subfigure}[b]{0.24\textwidth}
         \centering
\includegraphics[width=\textwidth]{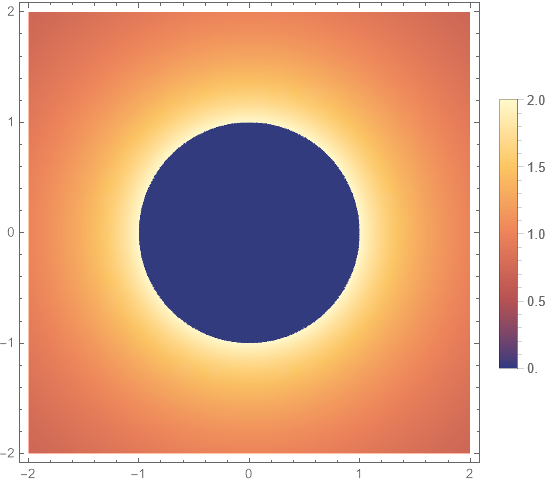}
         \caption{$\phi=0$}
         \label{fig:cvort0F}
     \end{subfigure}
     \hfill
     \begin{subfigure}[b]{0.24\textwidth}
         \centering
\includegraphics[width=\textwidth]{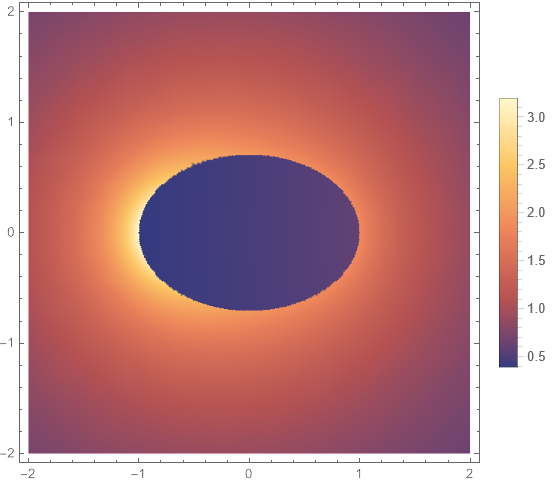}
         \caption{$\phi=\frac{\pi}{4}$}
         \label{fig:cvort45F}
     \end{subfigure}
     \hfill
     \begin{subfigure}[b]{0.24\textwidth}
         \centering
\includegraphics[width=\textwidth]{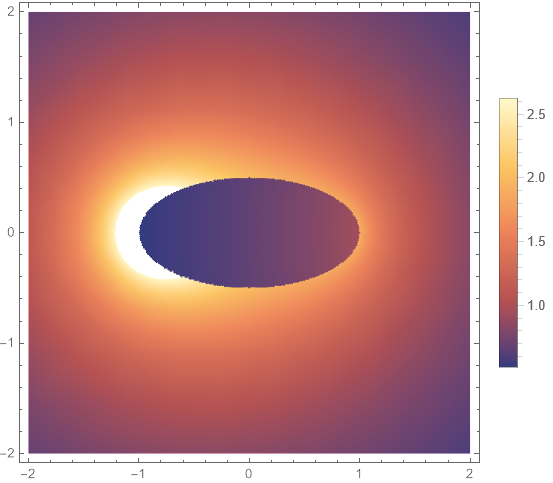}
         \caption{$\phi=\frac{\pi}{3}$}
         \label{fig:cvort60F}
     \end{subfigure}
     \hfill
     \begin{subfigure}[b]{0.24\textwidth}
         \centering
\includegraphics[width=\textwidth]{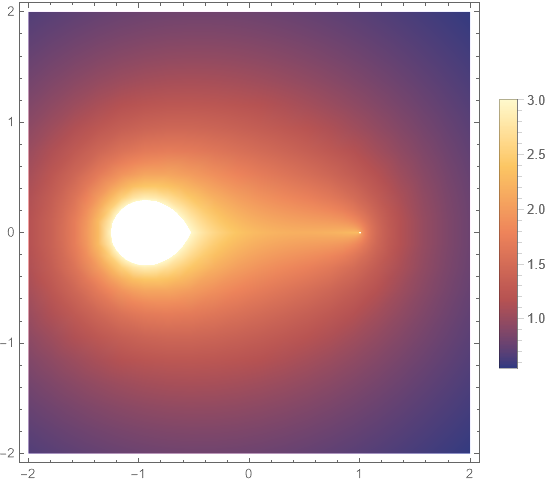}
         \caption{$\phi=\frac{\pi}{2}$}
         \label{fig:cvort90F}
     \end{subfigure}
        \caption{Deflection vector magnitude $|\Vec{F}|=\sqrt{F_1^2+F_2^2}$ from circular vorton of various $\phi$. Fig.~(d), for example, shows the deflection around the left side of the ring projection is the largest.}
        \label{fig:Fcvort}
\end{figure}

Here, we apply the formalism to the case of the circular vorton given by Eq.~\eqref{circvortonsol}. For comparison, we also apply it to the circular Nambu-Goto loop \eqref{circNGsol}, as both exhibit similar geometries. For the Nambu-Goto loop, we define the coordinate $\zeta\equiv R\theta$ and let $r_o(t)=\cos{t/R}$. At the time of maximum loop radius, $t_0=0$, we set $r_o(t_0)=1$. This set up leads to
\begin{eqnarray}
    F_1(x_1,x_2)&=&-\frac{1}{2\pi}\int_0^{2\pi}d\theta\frac{\left(1-\sin^2{\phi}\cos^2{\theta}\right)\left(\cos{\theta}-x_1\right)}{\left(\cos{\theta}-x_1\right)^2+\left(\cos{\phi}\sin{\theta}-x_2\right)^2},\nonumber\\
    F_2(x_1,x_2)&=&-\frac{1}{2\pi}\int_0^{2\pi}d\theta\frac{\left(1-\sin^2{\phi}\cos^2{\theta}\right)\left(\cos{\phi}\sin{\theta}-x_2\right)}{\left(\cos{\theta}-x_1\right)^2+\left(\cos{\phi}\sin{\theta}-x_2\right)^2}.
\end{eqnarray}
For the case of circular vorton, we define $\theta\equiv\left(t_0+\zeta\right)/2R$. It yields
\begin{eqnarray}
    F_1(x_1,x_2)&=&-\frac{1}{\pi}\int_0^{2\pi} d\theta \left[\frac{1-\sin{\phi}\cos{\theta}}{1-\frac{1}{2}\sin{\phi}\cos{\theta}}\frac{\cos{\theta}-x_1}{\left(\cos{\theta}-x_1\right)^2+\left(\cos{\phi}\sin{\theta}-x_2\right)^2}\right],\nonumber\\    
    F_2(x_1,x_2)&=&-\frac{1}{\pi}\int_0^{2\pi} d\theta \left[\frac{1-\sin{\phi}\cos{\theta}}{1-\frac{1}{2}\sin{\phi}\cos{\theta}}\frac{\cos{\phi}\sin{\theta}-x_2}{\left(\cos{\theta}-x_1\right)^2+\left(\cos{\phi}\sin{\theta}-x_2\right)^2}\right].    
\end{eqnarray}
In the perpendicular direction ($\phi=0$), both the Nambu-Goto loop and the circular vorton produce the same deflection vector $\vec{F}$; however, the magnitude for the circular vorton is twice that of the Nambu-Goto loop. %We can see that for the case of $\phi=0$, that is the perpendicular case, both the circular NG and vorton solution yields exactly the same deflection vector $\vec{F}$, though the magnitude of the vorton deflection vector (and each components) is twice that of the NG solution. 
By rescaling $C\rightarrow 2C$ or setting $\mu=2\mu$ and subsequently dividing the deflection vector by a factor of 2 (which leaves the lens equations unchanged), we obtain an identical deflection vectors for both cases. Thus, the gravitational lensing signatures of a circular Nambu-Goto string and a vorton with linear mass densities $\mu$ and $2\mu$, respectively, are identical when the loop plane is perpendicular to the optical axis. In this study, we set $C=2.250$ for the Nambu-Goto loop and $C=1.125$ for vortons, ensuring comparable conditions in the circular cases. The density plots in Figs.~\ref{fig:F12ccs}-\ref{fig:Fcvort} illustrates the components and magnitude of $\vec{F}$. Note that the coordinate runs from $(x_1,x_2)=(-2,-2)$ to $(x_1,x_2)=(2,2)$, with $x_1$ being the horizontal axis and $x_2$ the vertical axis. %, which is the case for all density plots in this section.%Thus, the gravitational lensing signature of circular NG string of linear mass density $\mu$ and vorton of linear mass density $2\mu$ are identical if the loop plane is perpendicular to the optical axis. Throughout this study, we will use the value $C=2.250$ for NG loop and $C=1.125$ for vortons, so that the circular cases are comparable. The density plot below shows the components and magnitude of $\Vec{F}$.

Figs.~\ref{fig:F12ccs}-\ref{fig:Fccs} show that the deflection is symmetric and uniform, in the sense that points near each part of the string projection experiencing similar levels of deflection. The positive and negative horizontal (vertical) deflection seen from $F_1$ ($F_2$) values in Fig.~\ref{fig:Fccs} are shown to be symmetric for the left side and right side of the projection. The plot in Fig.~\ref{fig:Fccs} confirms that the magnitude of the deflection vector itself exhibit left-right symmetry, and that the deflection is largest right around the outer side of the string projection, while the deflection around the inner part of the string projection is minimal. In contrast, the positive and negative horizontal (vertical) deflection by $F_1$ ($F_2$) in Figs.~\ref{fig:F12cvort}-\ref{fig:Fcvort} reveal that for the case of vorton, the positive horizontal deflection on the left side of the string projection is larger in magnitude (indicated by the divergence in computational value) than the magnitude of negative deflection on the right side of the string projection, which indicates left-right asymmetry in the deflection vector. This asymmetry is absent in the deflection vector of NG cosmic string, where the magnitude of positive and negative deflection to the left and right side is the same. The behavior of minimal deflection inside the string projection is retained. The results (both NG and vorton) show discontinuities at the string projection, defining the boundary between inner part of the projection and the outer part. This behavior arises from the residue theorem in the integral calculation. The asymmetry can be interpreted as behavior of the deflection near the string segments where the current is maximally aligned with the source are significantly stronger than in other regions. This phenomenon can be attributed to the fact that the photon has to undergo the frame dragging in the opposite direction of the light ray at those points. For $\phi=0$, both the Nambu-Goto loop and vorton yield identical results, as expected. The observed asymmetry, along with the absence of double imaging of the string, is noteworthy; it likely results from scale differences and the use of the thin-lens approximation.

\begin{figure}[H]
     \centering
     \begin{subfigure}[b]{0.24\textwidth}
         \centering
\includegraphics[width=\textwidth]{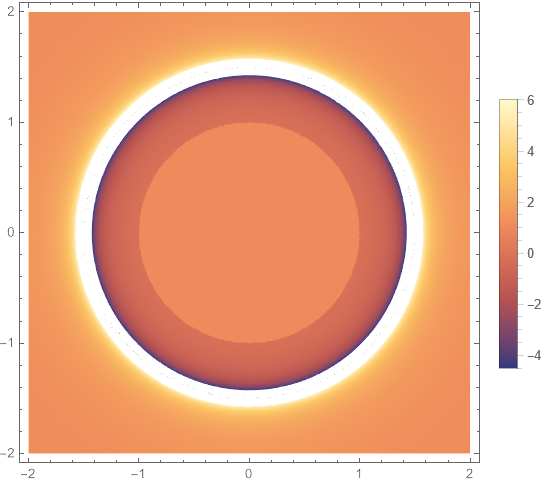}
         \caption{$\phi=0$}
         \label{fig:ccs0M}
     \end{subfigure}
     \hfill
     \begin{subfigure}[b]{0.24\textwidth}
         \centering
\includegraphics[width=\textwidth]{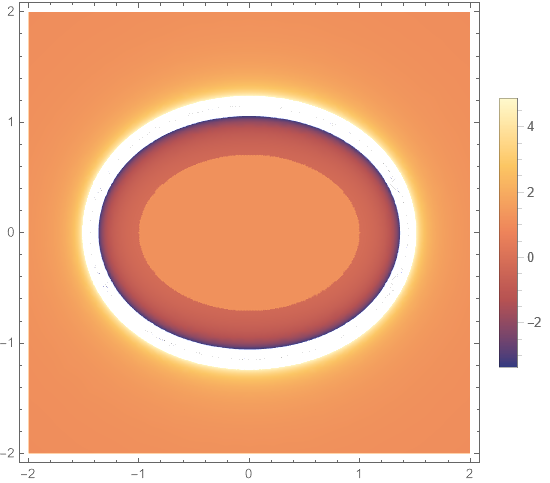}
         \caption{$\phi=\frac{\pi}{4}$}
         \label{fig:ccs45M}
     \end{subfigure}
     \hfill
     \begin{subfigure}[b]{0.24\textwidth}
         \centering
\includegraphics[width=\textwidth]{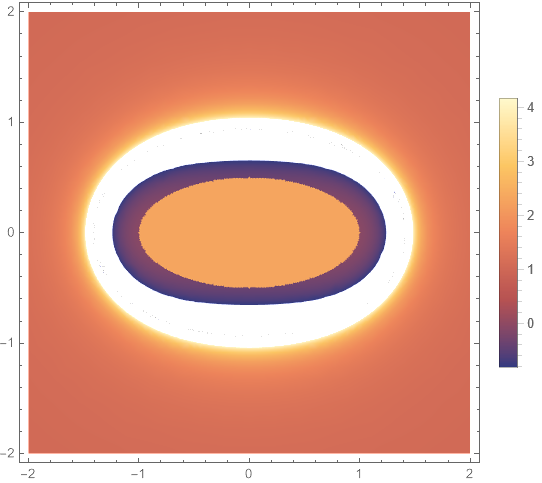}
         \caption{$\phi=\frac{\pi}{3}$}
         \label{fig:ccs60M}
     \end{subfigure}
     \hfill
     \begin{subfigure}[b]{0.24\textwidth}
         \centering
\includegraphics[width=\textwidth]{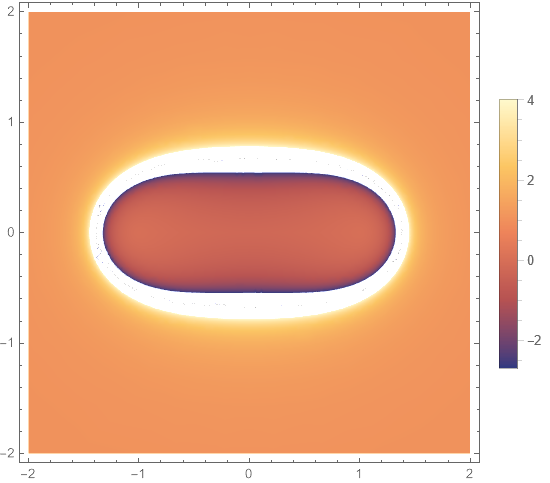}
         \caption{$\phi=\frac{\pi}{2}$}
         \label{fig:ccs90M}
     \end{subfigure}
        \caption{Magnification by circular Nambu-Goto loop for various $\phi$.}
        \label{fig:Mccs}
\end{figure}

\begin{figure}[H]
     \centering
     \begin{subfigure}[b]{0.24\textwidth}
         \centering
\includegraphics[width=\textwidth]{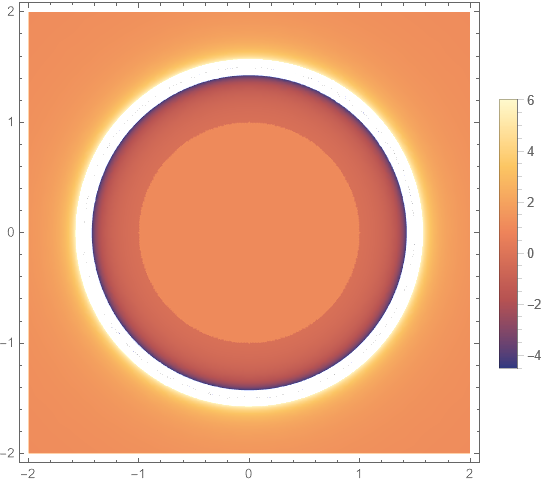}
         \caption{$\phi=0$}
         \label{fig:cvort0M}
     \end{subfigure}
     \hfill
     \begin{subfigure}[b]{0.24\textwidth}
         \centering
\includegraphics[width=\textwidth]{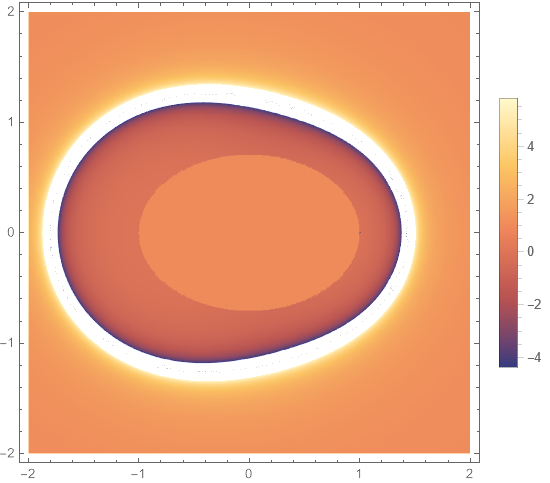}
         \caption{$\phi=\frac{\pi}{4}$}
         \label{fig:cvort45M}
     \end{subfigure}
     \hfill
     \begin{subfigure}[b]{0.24\textwidth}
         \centering
\includegraphics[width=\textwidth]{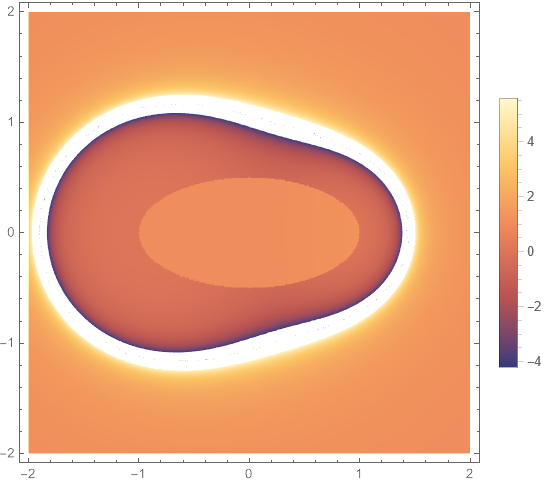}
         \caption{$\phi=\frac{\pi}{3}$}
         \label{fig:cvort60M}
     \end{subfigure}
     \hfill
     \begin{subfigure}[b]{0.24\textwidth}
         \centering
\includegraphics[width=\textwidth]{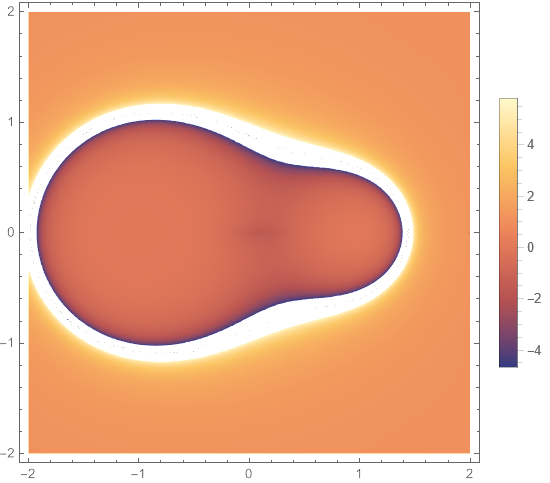}
         \caption{$\phi=\frac{\pi}{2}$}
         \label{fig:cvort90M}
     \end{subfigure}
        \caption{Magnification by circular vorton for various $\phi$.}
        \label{fig:Mcvort}
\end{figure}

Figs.~\ref{fig:Mccs}-\ref{fig:Mcvort} display density plots of the magnification for the circular Nambu-Goto loop and circular vorton at various values of $\phi$. Aside from the symmetry, it is notable that the magnification magnitude near the string is on the order of unity. We can also see that within the string projection, the magnification remains relatively constant and is also around unity. Due to the non-invertibility of the lens equations, there could be multiple images of the same source points. Consequently, an observer might see a relatively undistorted, source-like image ({\it e.g.,} of a galaxy) alongside more distorted images.

\begin{figure}[H]
     \centering
     \begin{subfigure}[b]{0.24\textwidth}
         \centering
\includegraphics[width=\textwidth]{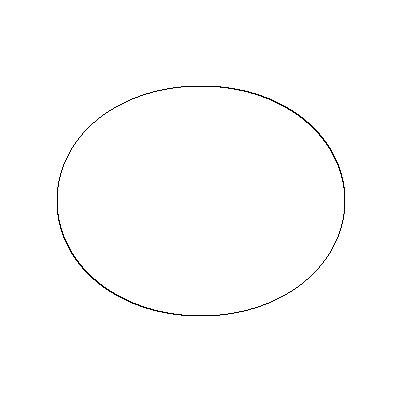}
         \caption{$\phi=\pi/4$}
\label{fig:critcurveccs45deg}
     \end{subfigure}
     \hfill
     \begin{subfigure}[b]{0.24\textwidth}
         \centering
\includegraphics[width=\textwidth]{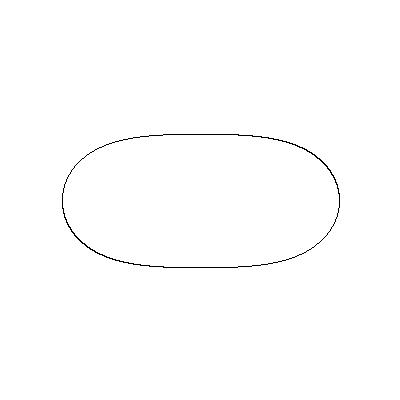}
         \caption{$\phi=\pi/2$}
    \label{fig:critcurveccs90deg}
     \end{subfigure}
     \hfill
     \begin{subfigure}[b]{0.24\textwidth}
         \centering
\includegraphics[width=\textwidth]{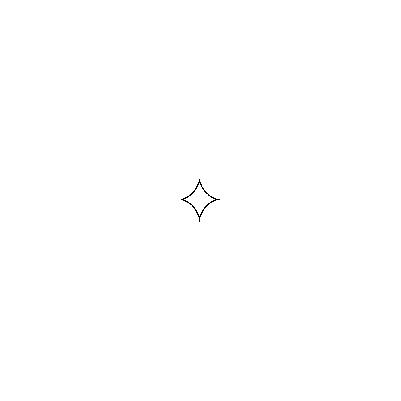}
         \caption{$\phi=\pi/4$}
\label{fig:ccausticccs45deg}
     \end{subfigure}
     \hfill
     \begin{subfigure}[b]{0.24\textwidth}
         \centering
\includegraphics[width=\textwidth]{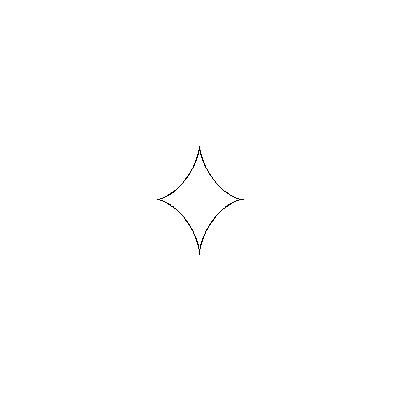}
         \caption{$\phi=\pi/2$}
\label{fig:ccausticccs90deg}
     \end{subfigure}
        \caption{Critical (a, b) and caustics (c, d) curves of the circular cosmic string of various $\phi$ at $t=0$. Both critical curves (the curve in the image plane which receives infinite magnification) and the caustics (the corresponding curves in the source plane) exhibit left-right and up-down symmetry. The shape in which the caustics and critical curve take is very much like those of curves from elliptic mass distribution sources, as seen in~\cite{Schneider:1992bmb}.}
\label{fig:ccscritcaustics}
\end{figure}

\begin{figure}[H]
     \centering
     \begin{subfigure}[b]{0.24\textwidth}
         \centering
\includegraphics[width=\textwidth]{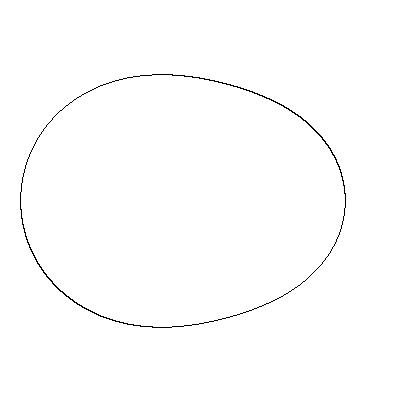}
         \caption{$\phi=\pi/4$}
\label{fig:critcurvevort45deg}
     \end{subfigure}
     \hfill
     \begin{subfigure}[b]{0.24\textwidth}
         \centering
\includegraphics[width=\textwidth]{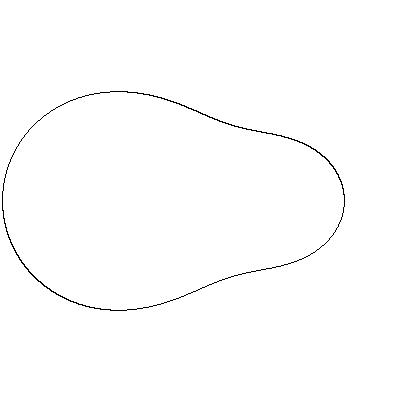}
         \caption{$\phi=\pi/2$}
    \label{fig:critcurvevort90deg}
     \end{subfigure}
     \hfill
     \begin{subfigure}[b]{0.24\textwidth}
         \centering
\includegraphics[width=\textwidth]{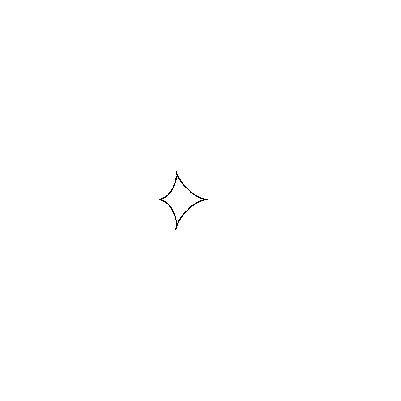}
         \caption{$\phi=\pi/4$}
    \label{fig:ccausticvort45deg}
     \end{subfigure}
     \hfill
     \begin{subfigure}[b]{0.24\textwidth}
         \centering
\includegraphics[width=\textwidth]{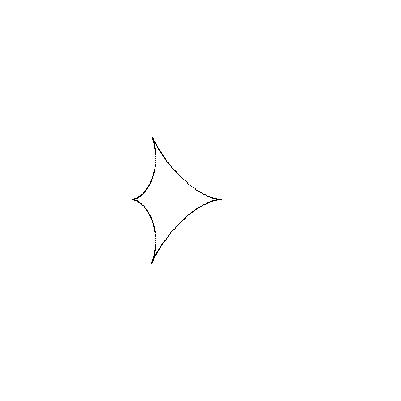}
         \caption{$\phi=\pi/2$}
    \label{fig:ccausticvort90deg}
     \end{subfigure}
        \caption{Critical (a, b) and caustics (c, d) curves of the circular vorton of various $\phi$. It can be seen that the vorton critical curves and caustics exhibits asymmetry, while maintaining similar shape with those of circular NG string.}
\label{fig:vircvortcritcaustics}
\end{figure}

The lensing images of the circular Nambu-Goto string and circular vorton can be effectively compared through their critical curves and caustics. As shown in Figs.~\ref{fig:ccscritcaustics} and \ref{fig:vircvortcritcaustics}, the critical curves and caustics of the circular Nambu-Goto string are symmetric while those of the circular vorton are asymmetric. This asymmetry arises due to the frame-dragging effect.

\begin{figure}[H]
     \centering
     \begin{subfigure}[b]{0.3\textwidth}
         \centering
\includegraphics[width=\textwidth]{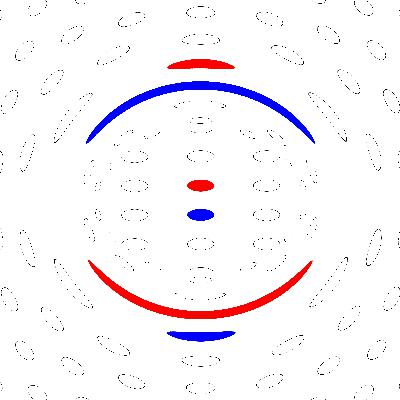}
         \caption{$\phi=\pi/4$}
    \label{fig:lensimccs45deg}
     \end{subfigure}
     \hfill
     \begin{subfigure}[b]{0.3\textwidth}
         \centering
\includegraphics[width=\textwidth]{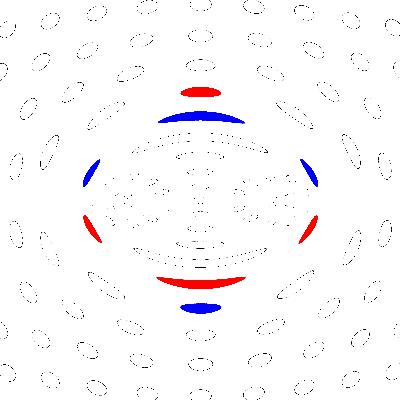}
         \caption{$\phi=\pi/2$}
    \label{fig:lensimccs90deg}
     \end{subfigure}
     \hfill
     \begin{subfigure}[b]{0.3\textwidth}
         \centering
\includegraphics[width=\textwidth]{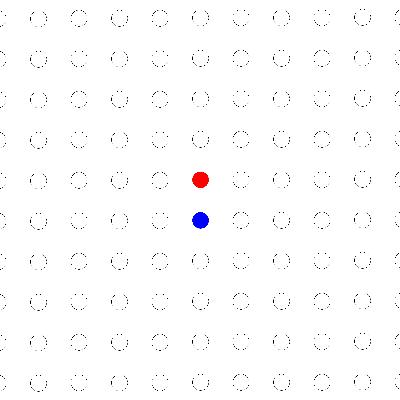}
         \caption{Unlensed}
    %\label{fig:lensimccs90deg}
 \end{subfigure}
\caption{Lensing Image of circular cosmic string loop at $\phi=\pi/4$ (a) and $\phi=\pi/2$ (b) and the unlensed image (c). In (a), the red and dot from source plane is depicted 3 times in the image plane. One of the image is an upright image inside the string projection, minimally distorted (though stretched horizontally), and two upright images right outside the string projection, one on top and one on the bottom. In (b), there are no inner string projection since $\phi=\pi/2$. Instead, there are two additional images for each red and blue circles, both upside down at the left and right side of the string projection. These are images inside the critical curves, while the images outside the critical curve is lensed upright.}
\label{fig:gridcircnsccs}
\end{figure}

\begin{figure}[H]
     \centering
     \begin{subfigure}[b]{0.3\textwidth}
         \centering
\includegraphics[width=\textwidth]{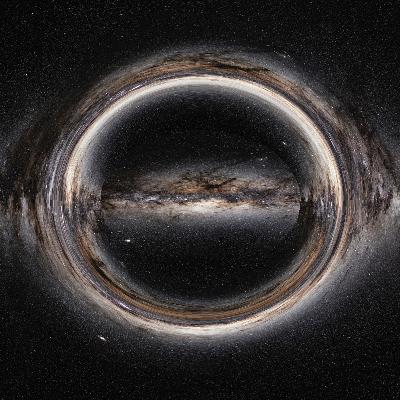}
         \caption{$\phi=\pi/4$}
    \label{fig:lensimccs45degmw}
     \end{subfigure}
     \hfill
     \begin{subfigure}[b]{0.3\textwidth}
         \centering
\includegraphics[width=\textwidth]{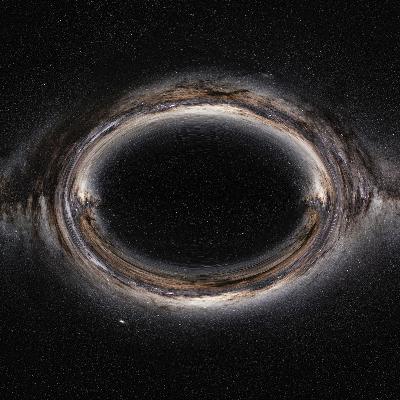}
         \caption{$\phi=\pi/2$}
    \label{fig:lensimccs90degmw}
     \end{subfigure}
     \hfill
     \begin{subfigure}[b]{0.3\textwidth}
         \centering
\includegraphics[width=\textwidth]{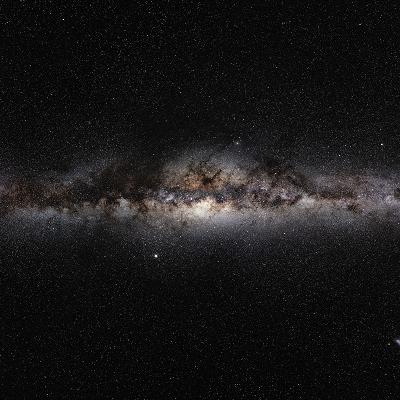}
         \caption{Unlensed}
    %\label{fig:lensimccs90deg}
     \end{subfigure}
\caption{Illustration of the Milky Way center lensed by circular cosmic string loop at $\phi=\pi/4$ (a), $\phi=\pi/2$ (b), and the unlensed image (c). The discontinuity in the image (a) is much clearer in this image than those in Fig.~\ref{fig:gridcircnsccs}.}
    \label{fig:mwccsn200}
\end{figure}

\begin{figure}[H]
     \centering
     \begin{subfigure}[b]{0.3\textwidth}
         \centering
\includegraphics[width=\textwidth]{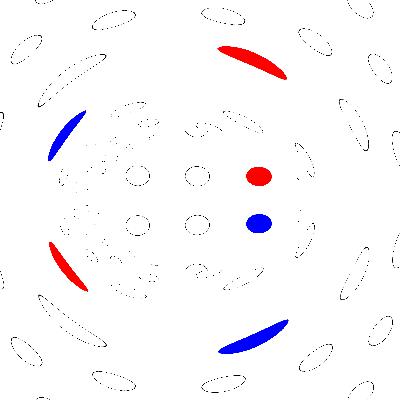}
         \caption{$\phi=\pi/4$}
    \label{fig:lensimcvort45deg}
     \end{subfigure}
     \hfill
     \begin{subfigure}[b]{0.3\textwidth}
         \centering
\includegraphics[width=\textwidth]{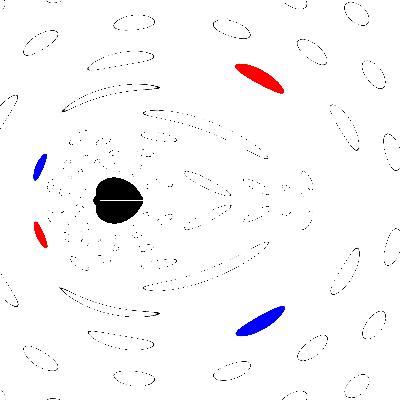}
         \caption{$\phi=\pi/2$}
    \label{fig:lensimcvort90deg}
     \end{subfigure}
        \caption{Lensing Image of circular vorton at $\phi=\pi/4$ (a) and $\phi=\pi/2$ (b). The source images is the same as those in Fig.~\ref{fig:gridcircnsccs}, and therefore the asymmetry in the images is readily apparent. In (a), the red and blue circle from the source image each appear three times. One upright image is inside the string projection, minimally distorted but shifted to the right. Another upright image is outside the string projection, with the red (blue) circle appears on the top (bottom) of the string projection. Another image is an upside-down image right to the left of the string projection. In (b), there is no inner part of the projection since $\phi=\pi/2$, and there are no extra images visible. The black area on the left side of the string is present because of the limited source image data that we use.}
    \label{fig:gridcircnscvort}
\end{figure}

\begin{figure}[H]
     \centering
     \begin{subfigure}[b]{0.3\textwidth}
         \centering
\includegraphics[width=\textwidth]{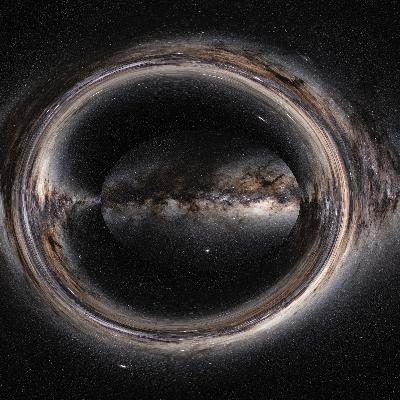}
         \caption{$\phi=\pi/4$}
    \label{fig:lensimcvort45degmw}
     \end{subfigure}
     \hfill
     \begin{subfigure}[b]{0.3\textwidth}
         \centering
\includegraphics[width=\textwidth]{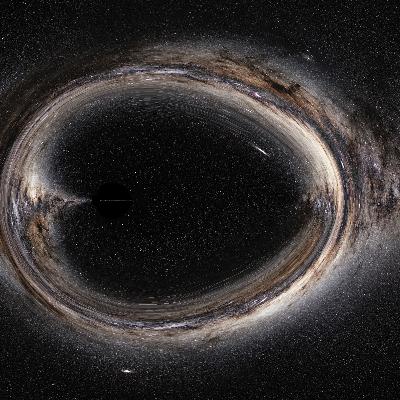}
         \caption{$\phi=\pi/2$}
    \label{fig:lensimcvort90degmw}
     \end{subfigure}
        \caption{Illustration of the Milky Way center lensed by circular vorton at $\phi=\pi/4$ (a) and $\phi=\pi/2$ (b). As before, the discontinuity in the image inside and outside the string projection is more apparent here than in Fig.~\ref{fig:gridcircnscvort}. The asymmetry of the image is also visible.}
        \label{fig:mwcvortn200}
\end{figure}

In Figs.~\ref{fig:gridcircnsccs}-\ref{fig:mwccsn200} and \ref{fig:gridcircnscvort}-\ref{fig:mwcvortn200} we can visibly observe a discontinuity between the images inside and outside the Nambu-Goto loop and the vorton, respectively. This is a generic feature in the thin-string approximation, which is expected to be smoothed out by the order of string's finite thickness $\delta$ or by using the full field equations. The difference of symmetry between the Nambu-Goto loop and the vorton case is more apparent, with the circular vorton producing an asymmetric Einstein ring despite its symmetric geometry. On the other hand, the Nambu-Goto loop yields a symmetric Einstein ring.

We can see that the region between the critical curve and the string is the region of inversion, where, as in Figs.~\ref{fig:gridcircnsccs} and \ref{fig:gridcircnscvort}, the blue circle takes place north of the red circle, where in the unlensed image it takes place south of the red circle, as it is the case of the image inside the string and outside the critical curve. The black region in Fig.~\ref{fig:lensimcvort90deg} (and several figures in the following discussions) is present because the image mapping algorithm ran out of pixel to sample from the source plane, in other words the $\Vec{y}$ output is outside the source image.

\subsection{Kibble-Turok Vorton}

Starting from the Kibble-Turok vorton in Eq.~\eqref{TurokVorton} and redefining $\theta\equiv \left(t_0+\zeta\right)/2R$, we obtain the deflection vector components %for the Kibble-Turok vorton with componentsgives the deflection vector of the Kibble-Turok vorton with components
\begin{eqnarray}
%\begin{split}
    F_1(x_1,x_2)&=&-\frac{1}{\pi}\int_0^{2\pi} d\theta \left(\frac{1-2\sqrt{\kappa(1-\kappa)}\sin{\theta}}{1-\sqrt{\kappa(1-\kappa)}\sin{\theta}}\right)
    \frac{(1-\kappa)\sin{\theta}+\frac{1}{3}\kappa \sin{(3\theta)}-x_1}{V},\nonumber\\     
%\end{split}   
%\end{equation}
%and
%\begin{equation}
%\begin{split}
    F_2(x_1,x_2)&=&\frac{1}{\pi}\int_0^{2\pi} d\theta \left(\frac{1-2\sqrt{\kappa(1-\kappa)}\sin{\theta}}{1-\sqrt{\kappa(1-\kappa)}\sin{\theta}}\right)
    \frac{(1-\kappa)\cos{\theta}+\frac{1}{3}\kappa\cos{(3\theta)}+x_2}{V},     %\end{split}   
\end{eqnarray}
where
\begin{equation}
    V\equiv\left((1-\kappa)\sin{\theta}+\frac{1}{3}\kappa \sin{(3\theta)}-x_1\right)^2+\left((1-\kappa)\cos{\theta}+\frac{1}{3}\kappa\cos{(3\theta)}+x_2\right)^2.
\end{equation}

%This means that an observer could see the (more or less) undistorted, source-like image of the source (a galaxy for example) together with the distorted images.

%\subsubsection{Critical Curves and Caustics}

\begin{figure}[H]
     \centering
     \begin{subfigure}[b]{0.3\textwidth}
         \centering
    \includegraphics[width=\textwidth]{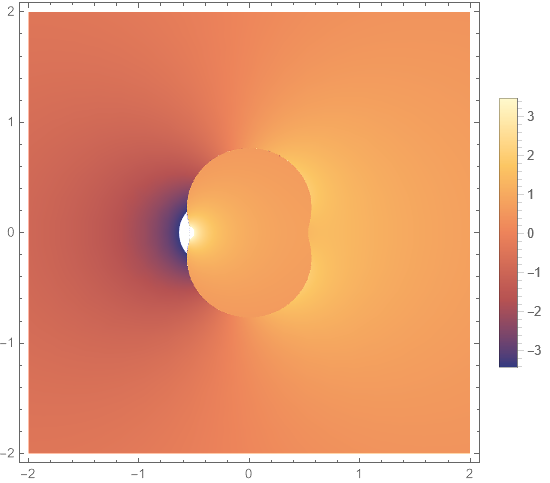}
         \caption{$\kappa=0.35$}
         \end{subfigure}
         \hfill
     \begin{subfigure}[b]{0.3\textwidth}
         \centering
    \includegraphics[width=\textwidth]{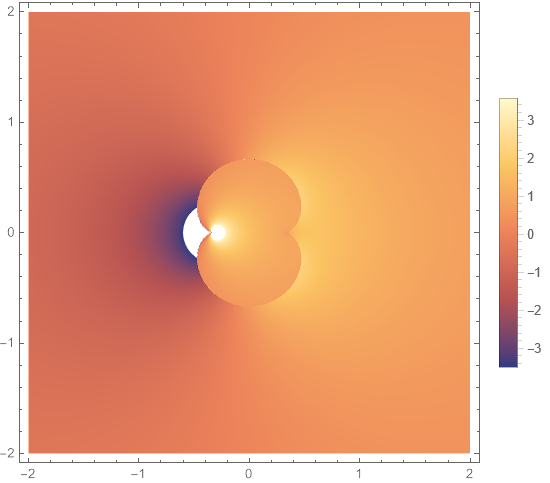}
         \caption{$\kappa=0.5$}
         \end{subfigure}
         \hfill
     \begin{subfigure}[b]{0.3\textwidth}
         \centering
    \includegraphics[width=\textwidth]{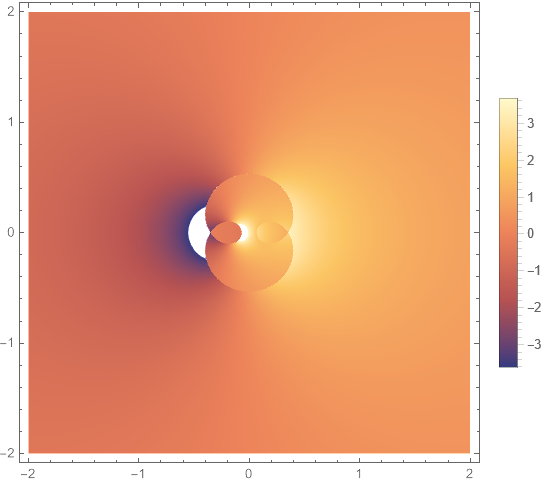}
         \caption{$\kappa=0.7$}
         \end{subfigure}
     \hfill
     \begin{subfigure}[b]{0.3\textwidth}
         \centering
    \includegraphics[width=\textwidth]{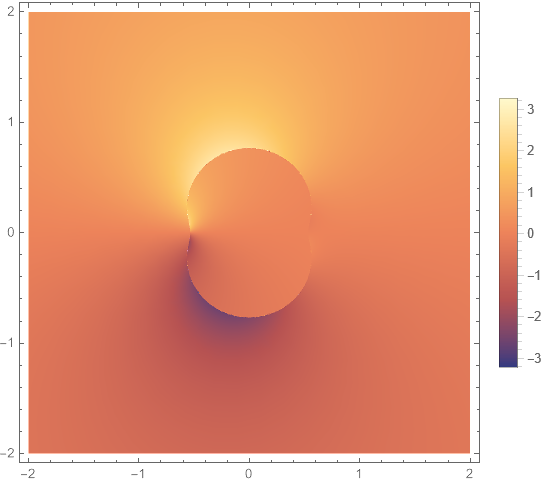}
         \caption{$\kappa=0.35$}
         \end{subfigure}
     \hfill
     \begin{subfigure}[b]{0.3\textwidth}
         \centering
    \includegraphics[width=\textwidth]{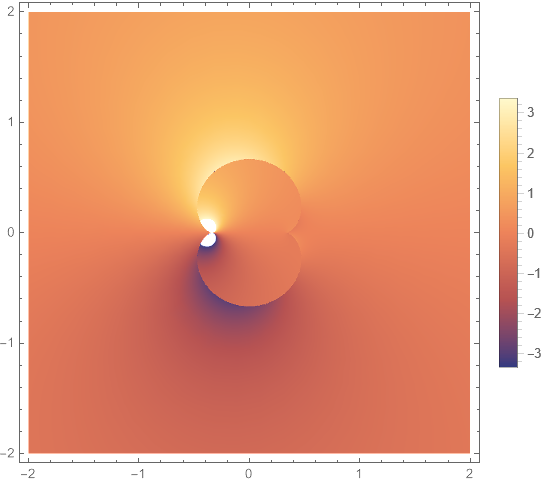}
         \caption{$\kappa=0.5$}
         \end{subfigure}
     \hfill
     \begin{subfigure}[b]{0.3\textwidth}
         \centering
    \includegraphics[width=\textwidth]{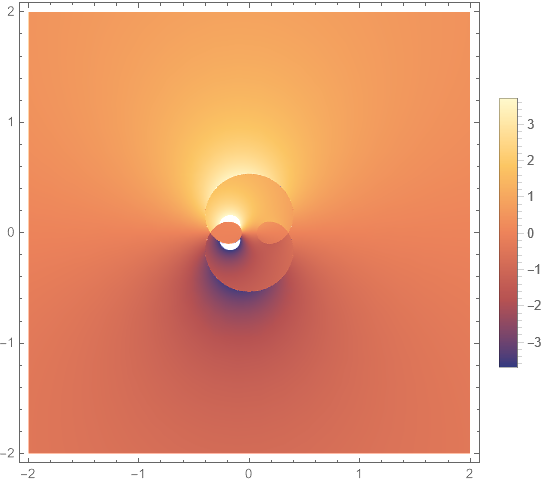}
         \caption{$\kappa=0.7$}
         \end{subfigure}
        \caption{Deflection vector component $F_1$ (a, b, c) and $F_2$ (d, e, f) from Kibble-Turok vorton of various $\kappa$. The asymmetry of the component is visible, as those seen in Fig.~\ref{fig:F12cvort}. In (c,f) there is a discontinuity in the image, not only on the boundary of the string projection but also in the intersection of string projection itself (it only intersects in projection, not in 3 dimension).}
        \label{fig:F12turokvort}
\end{figure}

\begin{figure}[H]
     \centering
     \begin{subfigure}[b]{0.3\textwidth}
         \centering
    \includegraphics[width=\textwidth]{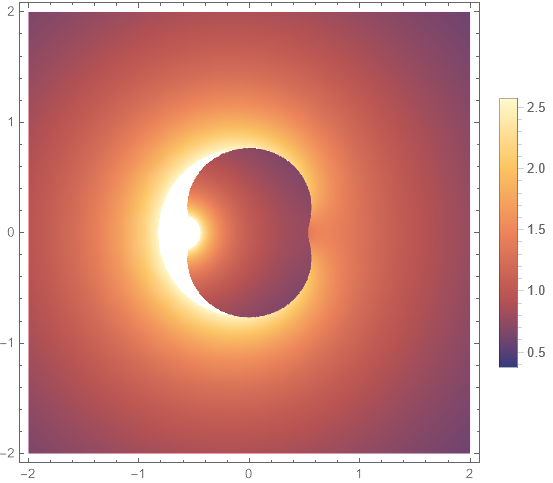}
         \caption{$\kappa=0.35$}
         \end{subfigure}
     \hfill
     \begin{subfigure}[b]{0.3\textwidth}
         \centering
    \includegraphics[width=\textwidth]{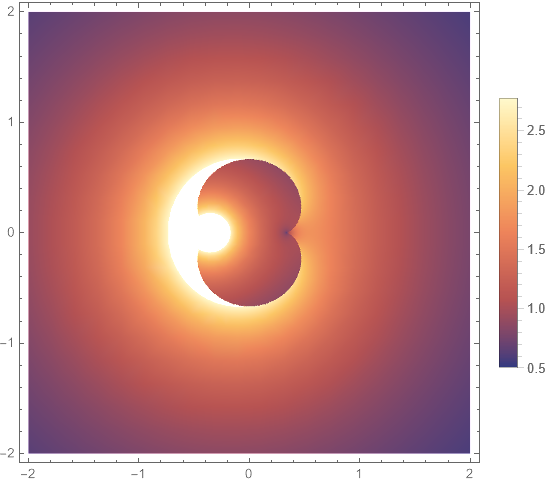}
         \caption{$\kappa=0.5$}
         \end{subfigure}
     \hfill
     \begin{subfigure}[b]{0.3\textwidth}
         \centering
    \includegraphics[width=\textwidth]{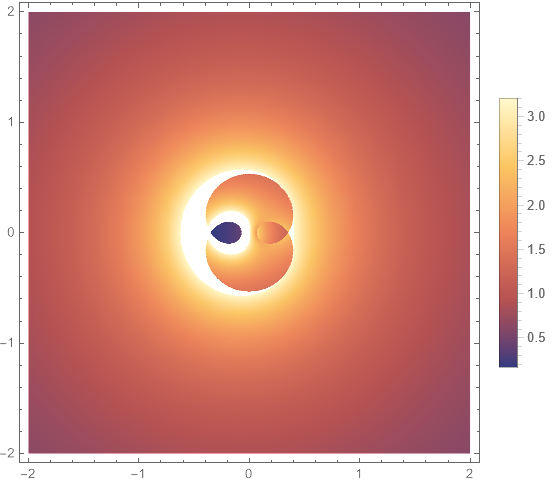}
         \caption{$\kappa=0.7$}
         \end{subfigure}
     \caption{Deflection vector magnitude $|\Vec{F}|=\sqrt{F_1^2+F_2^2}$ from Kibble-Turok vorton of various $\kappa$.}
        \label{fig:Fturokvort}
\end{figure}

The vector components and magnitude are shown as two-dimensional density plots in Figs.~\ref{fig:F12turokvort}-\ref{fig:Fturokvort}, respectively. As before, we observe a dominant deflection region despite the symmetry of the string projection. This comes from from the photon counteracting the frame-dragging effect, which acts in the opposite direction.

\begin{figure}[H]
     \centering
     \begin{subfigure}[b]{0.3\textwidth}
         \centering
    \includegraphics[width=\textwidth]{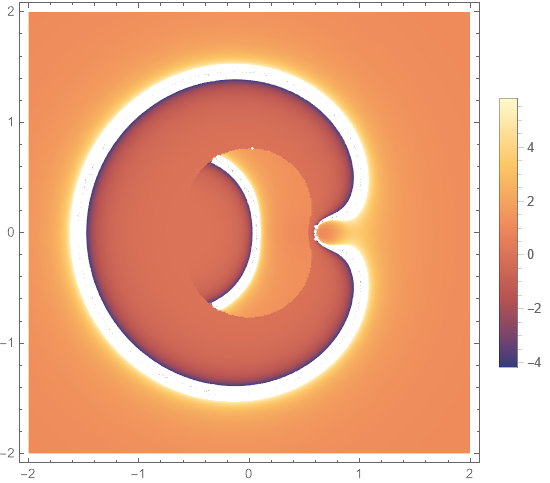}
         \caption{$\kappa=0.35$}
         \end{subfigure}
     \hfill
     \begin{subfigure}[b]{0.3\textwidth}
         \centering
    \includegraphics[width=\textwidth]{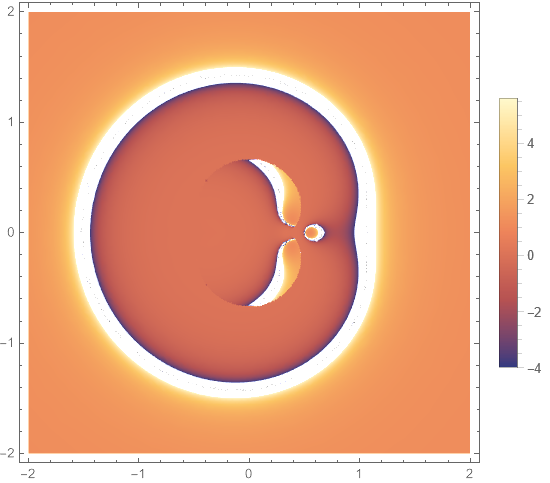}
         \caption{$\kappa=0.5$}
         \end{subfigure}
     \hfill
     \begin{subfigure}[b]{0.3\textwidth}
         \centering
    \includegraphics[width=\textwidth]{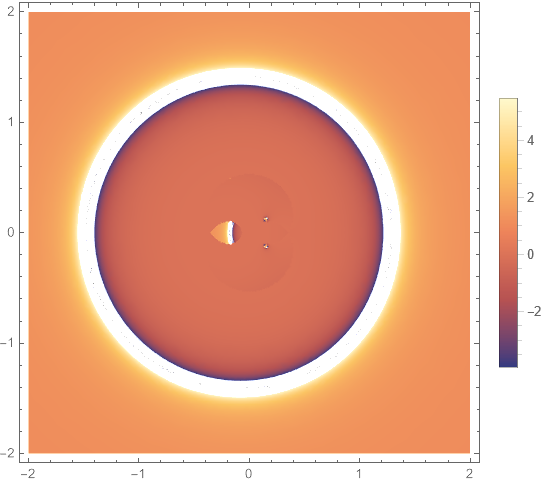}
         \caption{$\kappa=0.7$}
         \end{subfigure}
     \caption{Magnification by Kibble-Turok vorton for various $\kappa$.}
        \label{fig:Mturokvort}
\end{figure}

\begin{figure}[H]
     \centering
     \begin{subfigure}[b]{0.3\textwidth}
         \centering
    \includegraphics[width=\textwidth]{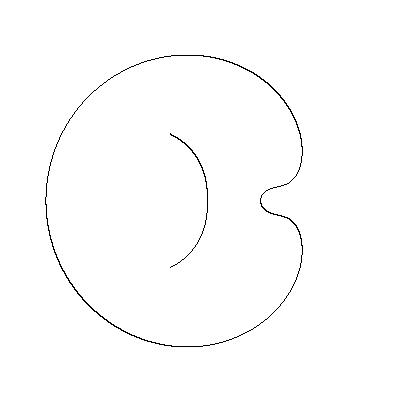}
         \caption{$\kappa=0.35$}
         \label{fig:critcurvevortk035}
     \end{subfigure}
     \hfill
     \begin{subfigure}[b]{0.3\textwidth}
         \centering
    \includegraphics[width=\textwidth]{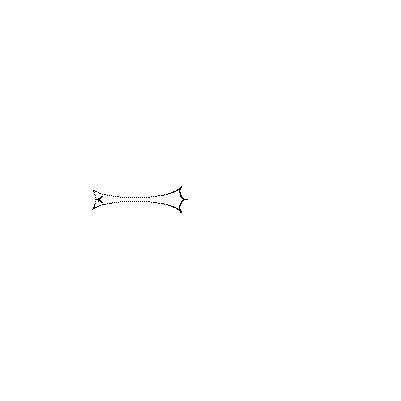}
         \caption{$\kappa=0.35$}
         \label{fig:causticvortk035}
     \end{subfigure}
     \hfill
     \begin{subfigure}[b]{0.3\textwidth}
         \centering
    \includegraphics[width=\textwidth]{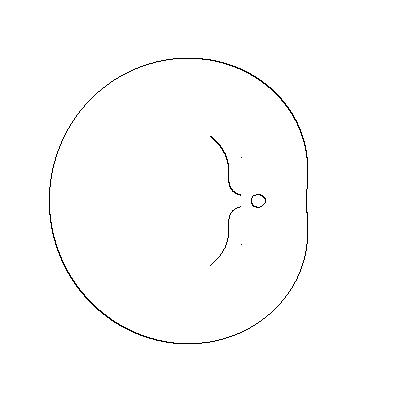}
         \caption{$\kappa=0.5$}
         \label{fig:critcurvevortk050}
     \end{subfigure}
     \hfill
     \begin{subfigure}[b]{0.3\textwidth}
         \centering
    \includegraphics[width=\textwidth]{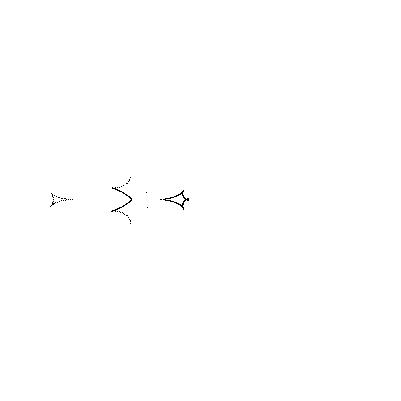}
         \caption{$\kappa=0.5$}
         \label{fig:causticvortk050}
     \end{subfigure}
     \hfill
     \begin{subfigure}[b]{0.3\textwidth}
         \centering
    \includegraphics[width=\textwidth]{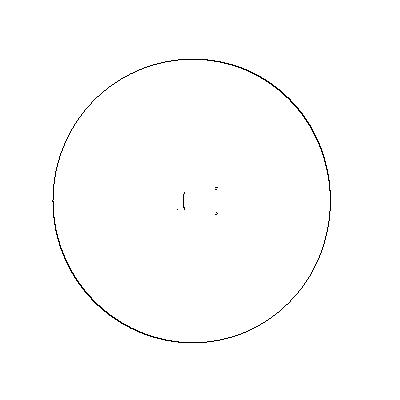}
         \caption{$\kappa=0.7$}
         \label{fig:critcurvevortk070}
     \end{subfigure}
     \hfill
     \begin{subfigure}[b]{0.3\textwidth}
         \centering
    \includegraphics[width=\textwidth]{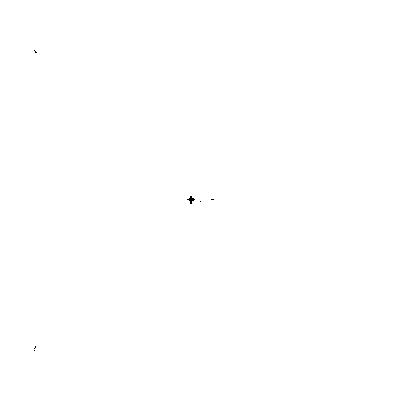}
         \caption{$\kappa=0.7$}
         \label{fig:causticvortk070}
     \end{subfigure}
        \caption{Critical (a, c, e) and caustics (b, d, f) curves of the Kibble-Turok vorton of various $\kappa$.}
    \label{fig:turokvortcritcaustics}
\end{figure}

The image magnifications from the Kibble-Turok vorton are shown in Fig.~\ref{fig:Mturokvort}. We can observe that there can be multiple regions of high magnification, where both inverted and non-inverted images inside the loop projection are visible. Fig.~\ref{fig:turokvortcritcaustics} displays the critical curves and caustics for several values of $\kappa$. Notably, multiple critical curves can be seen, with additional critical curve appearing within the string loop region. This discontinuity arises from the thin-string approximation and would be smoothed out if the approximation were relaxed.

%Finally, the lensing images are presented in Fig.~\ref{fig:gridcircnsturok}, while Fig.~\ref{fig:mwvortturokn200} illustrates the lensing effect on the Milky Way. As before, the frame-dragging effect produces visible asymmetry of the Einstein ring in \ref{fig:mwvortturokn200}, despite the symmetry of the projected string geometry. Note that the string loop is not fully contained within the lensing plane, as can be seen in Fig.~\ref{fig:turokvortonsol}.

%A comparison of the lensing images of the circular Nambu-Goto string and circular vorton can be made through their critical curves and caustics. It can be seen in 

%The lensing images of the circular Nambu-Goto string and circular vorton can be effectively compared through their critical curves and caustics. As shown in Figs.~\ref{fig:ccscritcaustics} and \ref{fig:vircvortcritcaustics}, the critical curves and caustics of the circular Nambu-Goto string are symmetric while those of the circular vorton are asymmetric. This asymmetry arises due to the frame-dragging effect.

%We can compare the lensing images of the circular NG string and circular vorton from their critical curves and caustics. It can be seen clearly that the critical curve and caustics of the circular NG string, Fig.~\ref{fig:ccscritcaustics}, is symmetric, while the critical curves and caustics of circular vorton are asymmetric, due to the frame dragging effect previously discussed.
%\subsubsection{Lensing Image}
%The lensing images of circular NG string and vorton are as follows.
\begin{figure}[H]
     \centering
     \begin{subfigure}[b]{0.32\textwidth}
         \centering
    \includegraphics[width=\textwidth]{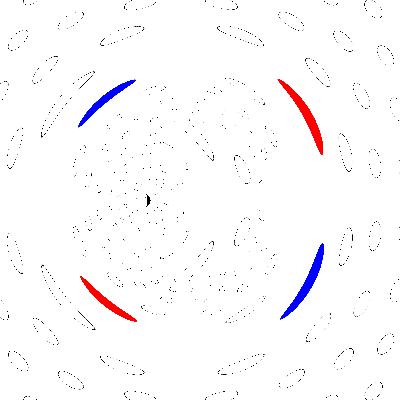}
         \caption{$\kappa=0.35$}
         \label{fig:lensimcvortk035}
     \end{subfigure}
     \hfill
     \begin{subfigure}[b]{0.3\textwidth}
         \centering
    \includegraphics[width=\textwidth]{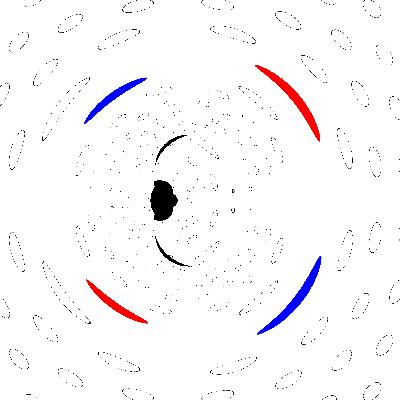}
         \caption{$\kappa=0.5$}
         \label{fig:lensimcvortk050}
     \end{subfigure}
     \hfill
     \begin{subfigure}[b]{0.3\textwidth}
         \centering
    \includegraphics[width=\textwidth]{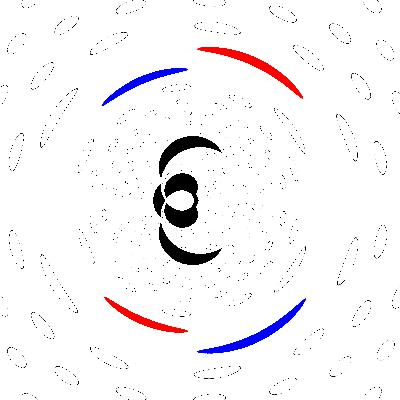}
         \caption{$\kappa=0.7$}
         \label{fig:lensimcvortk070}
     \end{subfigure}
        \caption{Lensing Image of Turok vorton for various values of $\kappa$.}
        \label{fig:gridcircnsturok}
\end{figure}

\begin{figure}[H]
     \centering
     \begin{subfigure}[b]{0.3\textwidth}
         \centering
    \includegraphics[width=\textwidth]{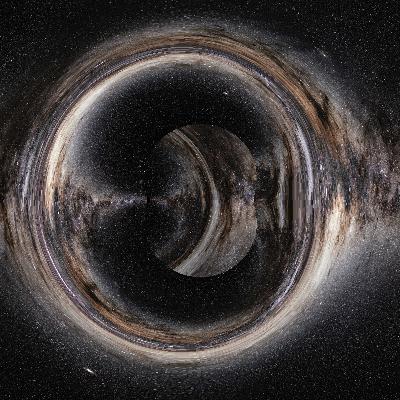}
         \caption{$\kappa=0.35$}
         \label{fig:lensimcvortk035mw}
     \end{subfigure}
     \hfill
     \begin{subfigure}[b]{0.3\textwidth}
         \centering
    \includegraphics[width=\textwidth]{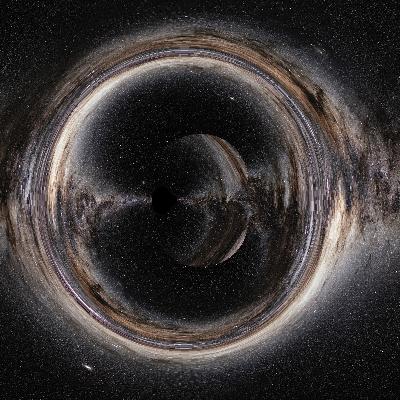}
         \caption{$\kappa=0.5$}
         \label{fig:lensimcvortk050mw}
     \end{subfigure}
     \hfill
     \begin{subfigure}[b]{0.3\textwidth}
         \centering
    \includegraphics[width=\textwidth]{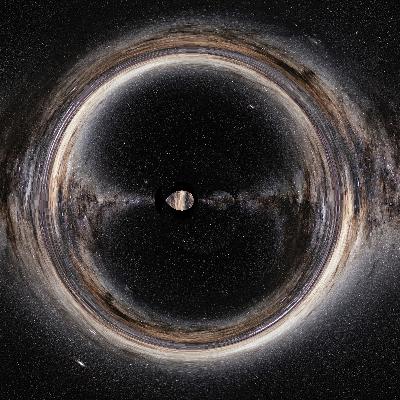}
         \caption{$\kappa=0.7$}
         \label{fig:lensimcvortk070mw}
     \end{subfigure}
        \caption{Illustration of the Milky Way center lensed by Turok vorton for various values of $\kappa$. It is apparent from (a) that an Einstein ring could form partially inside the string projection for this string solution. The discontinuity of image inside the intersection of string projection in (c) is notable.}
        \label{fig:mwvortturokn200}
\end{figure}

Finally, the lensing images are presented in Fig.~\ref{fig:gridcircnsturok}, while Fig.~\ref{fig:mwvortturokn200} illustrates the lensing effect on the Milky Way. As before, the frame-dragging effect produces visible asymmetry of the Einstein ring in \ref{fig:mwvortturokn200}, despite the symmetry of the projected string geometry. Note that the string loop is not fully contained within the lensing plane, as can be seen in Fig.~\ref{fig:turokvortonsol}.
%In Figs.~\ref{fig:gridcircnsccs}-\ref{fig:mwccsn200} and \ref{fig:gridcircnscvort}-\ref{fig:mwcvortn200} we can visibly observe a discontinuity between the images inside and outside the Nambu-Goto loop and the vorton, respectively. This is a generic feature in the thin-string approximation, which is expected to be smoothed out by the order of string's finite thickness $\delta$ or by using the full field equations. The difference of symmetry between the Nambu-Goto loop and the vorton case is more apparent, with the circular vorton producing an asymmetric Einstein ring despite its symmetric geometry. On the other hand, the Nambu-Goto loop yields a symmetric Einstein ring. %, reflecting the symmetry in both its geometry and lensing features. %where we could observe an asymmetric Einstein ring in the case of circular vorton, even though the geometry of the string itself is symmetric, unlike the case of circular NG string, where both the geometry and the Einstein ring are symmetric.
%As before, we could see the dominant part of the deflection even though the string projection is symmetric, where the dominant part of the deflection is due to the photon having to counter the frame dragging effect working in the opposite direction.

%\subsubsection{Magnification}
%Here are the magnification of images from Kibble-Turok vorton.

\subsection{The 123-Vorton}

Using the same re-definition for $\theta$ as before gives us the deflection vectors of the $123$-vorton 
\begin{equation}
%\begin{split}
    F_1(x_1,x_2)=-\frac{1}{\pi}\int_0^{2\pi} d\theta \left[\left(1-\frac{W}{1-W}\right)\frac{\sqrt{\beta}(1-\kappa)\sin{\theta}+\frac{1}{3}\kappa \sin{3\theta}-x_1}{\left(\sqrt{\beta}(1-\kappa)\sin{\theta}+\frac{1}{3}\kappa \sin{3\theta}-x_1\right)^2+\mathcal{R}_2^2}\right],     
%\end{split}   
\end{equation}
\begin{equation}
\begin{split}
    F_2(x_1,x_2)&=\frac{1}{\pi}\int_0^{2\pi} d\theta \left[\left(1-\frac{W}{1-W}\right)\right.\\
    &\left.\times\frac{\sqrt{\beta}(1-\kappa)\cos{\theta}-\frac{1}{\sqrt{2}}\sqrt{1-2\beta-\kappa^2}\sin{2\theta}-\sqrt{\frac{\beta}{2}}\cos{2\theta}+\frac{1}{3}\kappa\cos{3\theta}+x_2}{\left(\sqrt{\beta}(1-\kappa)\sin{\theta}+\frac{1}{3}\kappa \sin{3\theta}-x_1\right)^2+\mathcal{R}_2^2}\right],     
\end{split}   
\end{equation}
with
\begin{eqnarray}
W&\equiv&\frac{1-2\left(\sqrt{\beta}\sqrt{\kappa(1-\kappa)}\sin{\theta}-\sqrt{\beta/2}\kappa\sin{2\theta}\right)}{1-\left(\sqrt{\beta}\sqrt{\kappa(1-\kappa)}\sin{\theta}-\sqrt{\beta/2}\kappa\sin{2\theta}\right)},\nonumber\\
\mathcal{R}_2&\equiv&\sqrt{\beta}(1-\kappa)\cos{\theta}-\frac{1}{\sqrt{2}}\sqrt{1-2\beta-\kappa^2}\sin{2\theta}-\sqrt{\frac{\beta}{2}}\cos{2\theta}+\frac{1}{3}\kappa\cos{3\theta}+x_2.
\end{eqnarray}

%Here we can observe the existence of more than one critical curves, where another critical curve can reside inside the string loop region. Again, this discontinuity is the result of thin string approximation, and will be smoothed out once the approximation is lifted.
%\subsubsection{Lensing Image}
%Here are the lensing images from Kibble-Turok vorton for several values of $\kappa$.
\begin{figure}[H]
     \centering
     \begin{subfigure}[b]{0.24\textwidth}
         \centering
    \includegraphics[width=\textwidth]{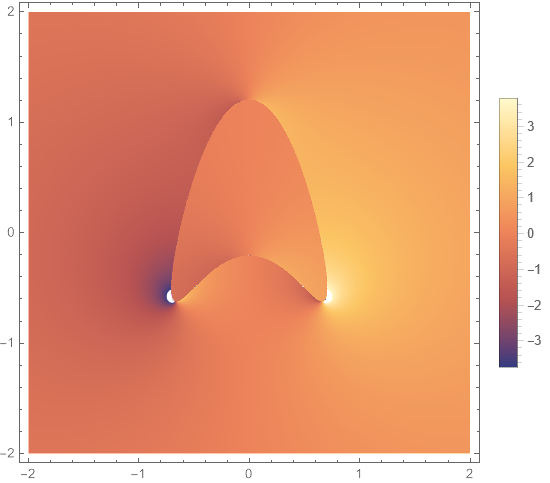}
         \caption{$\kappa=0$, $\beta=0.5$}
         \end{subfigure}
         \hfill
     \begin{subfigure}[b]{0.24\textwidth}
         \centering
    \includegraphics[width=\textwidth]{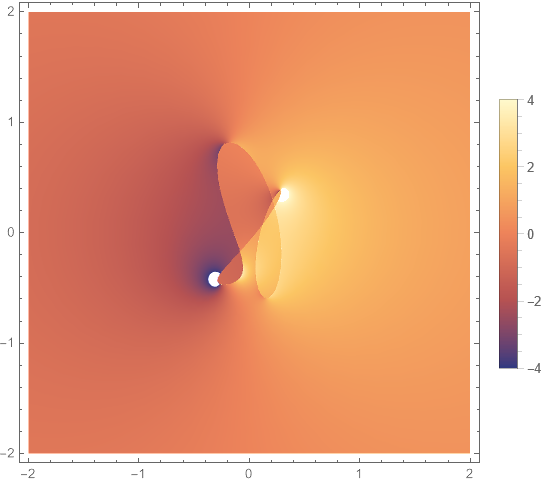}
         \caption{$\kappa=0.5$, $\beta=0.2$}
         \end{subfigure}
     \hfill
     \begin{subfigure}[b]{0.24\textwidth}
         \centering
    \includegraphics[width=\textwidth]{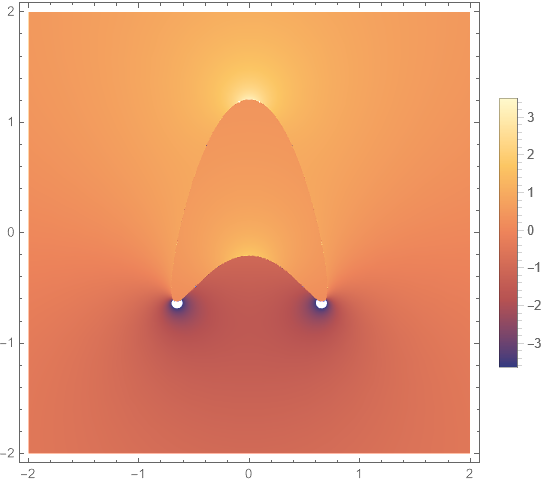}
         \caption{$\kappa=0$, $\beta=0.5$}
         \end{subfigure}
     \hfill
     \begin{subfigure}[b]{0.24\textwidth}
         \centering
    \includegraphics[width=\textwidth]{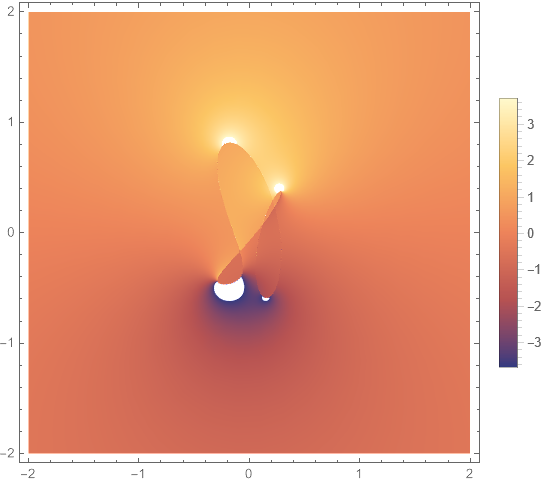}
         \caption{$\kappa=0.5$, $\beta=0.2$}
         \end{subfigure}
        \caption{Deflection vector component $F_1$ (a, b) and $F_2$ (c, d) of the $123$ vorton of various $\kappa$ and $\beta$. There is an apparent left-right symmetry in (a) and (c).}
        \label{fig:F12123vort}
\end{figure}
\begin{figure}[H]
     \centering
     \begin{subfigure}[b]{0.3\textwidth}
         \centering
    \includegraphics[width=\textwidth]{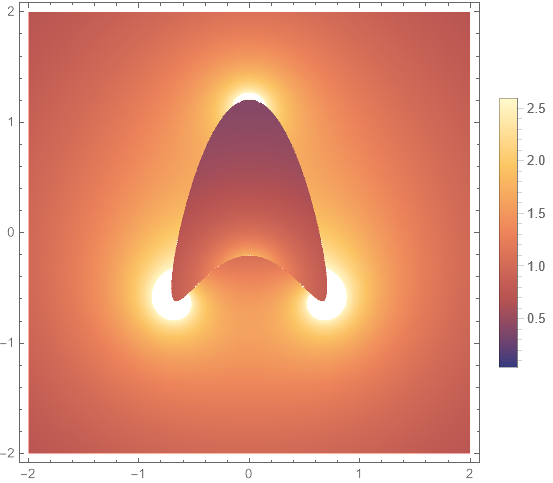}
         \caption{$\kappa=0$, $\beta=0.5$}
         \end{subfigure}
     \hfill
     \begin{subfigure}[b]{0.3\textwidth}
         \centering
    \includegraphics[width=\textwidth]{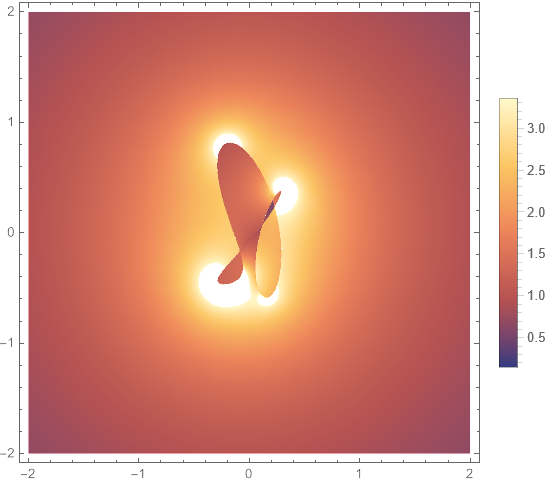}
         \caption{$\kappa=0.5$, $\beta=0.2$}
         \end{subfigure}
        \caption{Deflection vector magnitude $|\Vec{F}|=\sqrt{F_1^2+F_2^2}$ of the $123$ vorton of various $\kappa$ and $\beta$.}
        \label{fig:F123vort}
\end{figure}

The components and magnitudes of the vector are shown represented as density plots in Figs.~\ref{fig:F12123vort} and~\ref{fig:F123vort}, respectively. Unlike the cases before, the projected curve of the string of $\kappa=0$, $\beta=0.5$ is the actual vorton curve, as it lies completely on the lens plane. As the result, the effect of frame dragging would be irrelevant in this case, and hence the apparent symmetry of the deflection. However, we could still see some peaks of deflection on the image. This is not due to the frame dragging effect (as is the case in the $\kappa=0.5$, $\beta=0.2$ case), but the result of the locally peaked curvature of the string, which yields locally high energy density. This effect can also be obseved in the $\kappa=0.5$, $\beta=0.2$ case, however it is not present in the circular loop case, because the highly curved segments of the projected curve are only an apparent one, as the true shape of the loop is circular, which has uniform curvature.

\begin{figure}[H]
     \centering
     \begin{subfigure}[b]{0.3\textwidth}
         \centering
    \includegraphics[width=\textwidth]{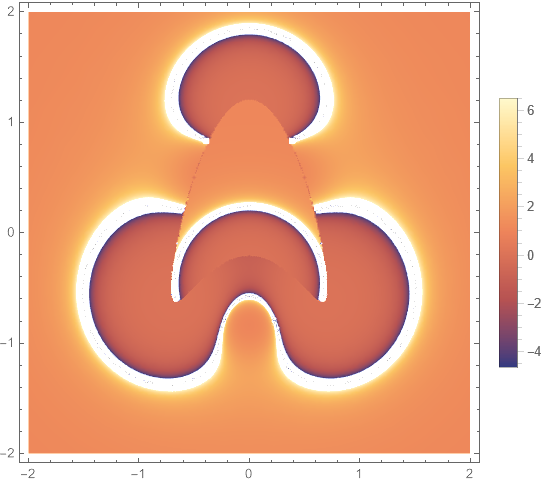}
         \caption{$\kappa=0$, $\beta=0.5$}
         \end{subfigure}
     \hfill
     \begin{subfigure}[b]{0.3\textwidth}
         \centering
    \includegraphics[width=\textwidth]{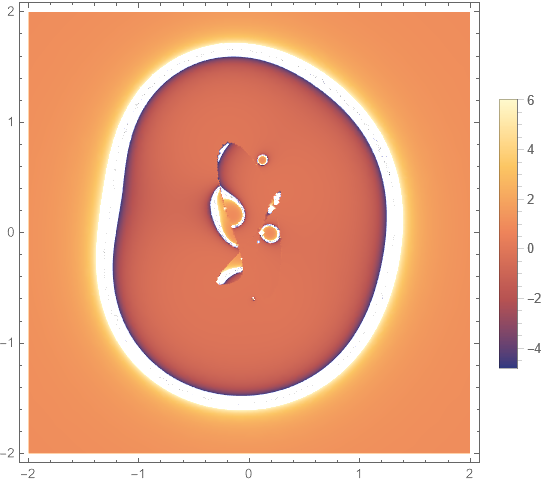}
         \caption{$\kappa=0.5$, $\beta=0.2$}
         \end{subfigure}
        \caption{Magnification by the $123$ vorton for several values of $\kappa$ and $\beta$.}
        \label{fig:M123vort}
\end{figure}

\begin{figure}[H]
     \centering
     \begin{subfigure}[b]{0.24\textwidth}
         \centering
    \includegraphics[width=\textwidth]{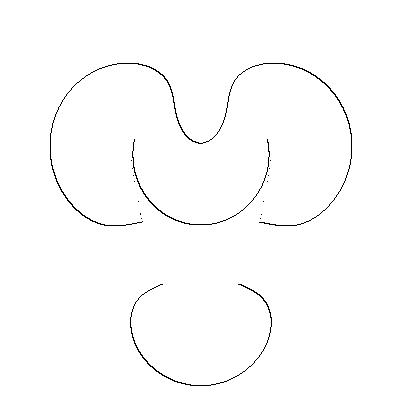}
         \caption{$\kappa=0$, $\beta=0.5$}
        \label{fig:critcurvevortk0b05}
     \end{subfigure}
     \hfill
     \begin{subfigure}[b]{0.24\textwidth}
         \centering
    \includegraphics[width=\textwidth]{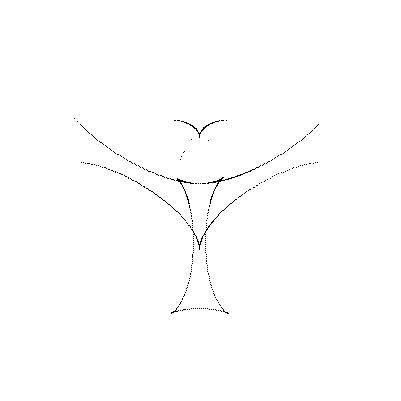}
         \caption{$\kappa=0$, $\beta=0.5$}
         \label{fig:causticvortk0b05}
     \end{subfigure}
     \hfill
     \begin{subfigure}[b]{0.24\textwidth}
         \centering
    \includegraphics[width=\textwidth]{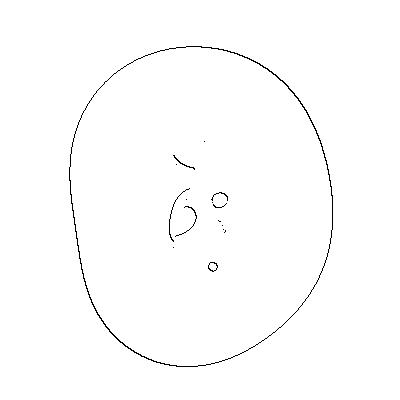}
         \caption{$\kappa=0.5$, $\beta=0.2$}
    \label{fig:critcurvevortk05b02}
     \end{subfigure}
     \hfill
     \begin{subfigure}[b]{0.24\textwidth}
         \centering
    \includegraphics[width=\textwidth]{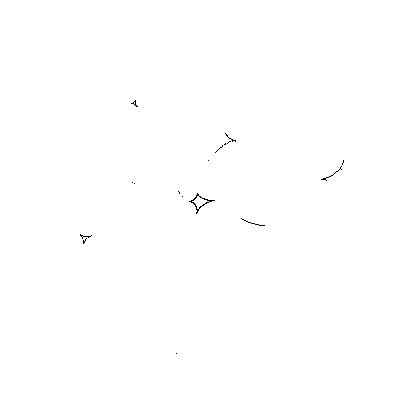}
         \caption{$\kappa=0.5$, $\beta=0.2$}
         \label{fig:causticvortk05b02}
     \end{subfigure}
        \caption{Critical (a, c) and caustics (b, d) curves of the $123$ vorton of various $\kappa$ and $\beta$.}
    \label{fig:123vortcritcaustics}
\end{figure}

The magnification density plots are displayed in Fig.~\ref{fig:M123vort}. As in previous cases, there are multiple regions with both inverted and non-inverted images, observable within the discontinuity of the loop's projected curve. The corresponding critical and caustics curves are shown in Fig.~\ref{fig:123vortcritcaustics}, where they appear perfectly symmetric, as does the string projection in Fig.~\ref{fig:2dk0b05} also being symmetric. This is because the vorton loop is perfectly inside the lensing plane. Consequently, there is no frame-dragging effect around an axis perpendicular to the optical axis, and the frame-dragging effect around the optical axis itself is negligible within this approximation. %, and therefore the frame dragging around the axis perpendicular to the optical axis does not exist, and the frame dragging effect around the optical axis itself is negligible in this approximation. % As before, there are multiple region of inverted image and non-inverted images, that can be observed inside the discontinuity of the loop projected curve.

\begin{figure}[H]
     \centering
     \begin{subfigure}[b]{0.3\textwidth}
         \centering
    \includegraphics[width=\textwidth]{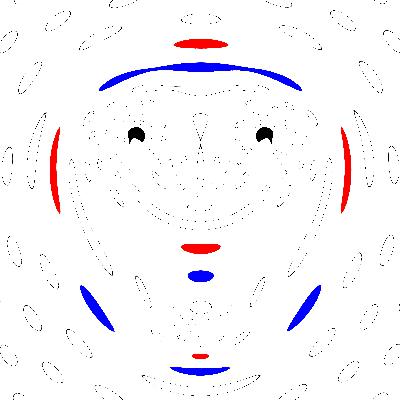}
         \caption{$\kappa=0$, $\beta=0.5$}
         \label{fig:lensimcvortk0b05}
     \end{subfigure}
     \hfill
     \begin{subfigure}[b]{0.3\textwidth}
         \centering
    \includegraphics[width=\textwidth]{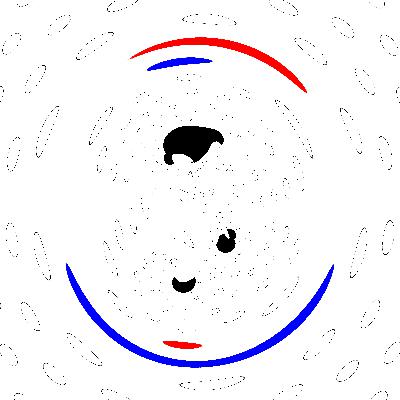}
         \caption{$\kappa=0.5$, $\beta=0.2$}
         \label{fig:lensimcvortk05b02}
     \end{subfigure}
        \caption{Lensing images of the $123$ vorton for various values of $\kappa$ and $\beta$.}
        \label{fig:gridcircns123}
\end{figure}

The lensing images along with its simulated effect on the Milky Way are shown in Figs.~\ref{fig:gridcircns123} and \ref{fig:mwvort123}, respectively, for several different values of $\kappa$ and $\beta$. As previously discussed, lensing effects such as the Einstein ring are evident in Fig.~\ref{fig:lensimcvortk0b05mw}, along with the pronounced symmetry visible in Fig.~\ref{fig:gridcircns123}. The discontinuity on the string, projected onto the image plane, is also clearly apparent. Other generic properties of lensing images, such as shear, is also present consistently across all images produced by all types of vorton and cosmic string discussed here. However, we note that in order to observe these discontinuities, the vorton ought to have at least a galaxy behind it, and that the vorton is of the angular size resolvable to telescope.

%We can see that unlike the vorton cases before, the critical curves \ref{fig:critcurvevortk0b05} and caustics \ref{fig:causticvortk0b05} are perfectly symmetric with the string projection \ref{fig:2dk0b05} also being symmetric. This is because unlike the previous cases, and like the case of $\phi=0$ in circular case, the vorton loop is perfectly inside the lensing plane, and therefore the frame dragging around the axis perpendicular to the optical axis does not exist, and the frame dragging effect around the optical axis itself is negligible in this approximation.

%\subsubsection{Lensing Image}

%Below are the lensing images of the $123$ vorton for different values of $\kappa$ and $\beta$.

\section{Conclusions}

In our previous work~\cite{Putra:2024zms}, we analyzed the lensing patterns of circular chiral vortons by deriving their metric and studying the associated null geodesics. In this study, we have extended the analysis to non-circular configurations. Specifically, we derived stationary solutions to the Nambu–Goto equations describing vorton loops with arbitrary harmonic modes and investigated their gravitational lensing effects. While this generalization sacrifices an explicit metric solution, we adopted the thin-lens approximation, treating the vorton’s contribution to the spacetime geometry as a weak perturbation on a flat background~\cite{deLaix:1996vc}.

\begin{figure}[H]
     \centering
     \begin{subfigure}[b]{0.3\textwidth}
         \centering
    \includegraphics[width=\textwidth]{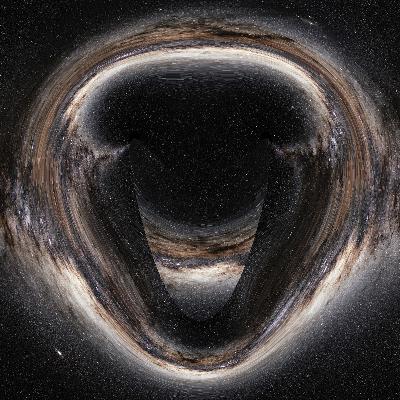}
         \caption{$\kappa=0$, $\beta=0.5$}
        \label{fig:lensimcvortk0b05mw}
     \end{subfigure}
     \hfill
     \begin{subfigure}[b]{0.3\textwidth}
         \centering   \includegraphics[width=\textwidth]{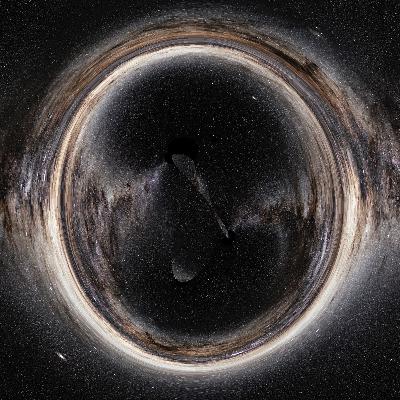}
    \caption{$\kappa=0.5$, $\beta=0.2$}
    \label{fig:lensimcvortk05b02mw}
     \end{subfigure}
        \caption{Illustration of the Milky Way center lensed by the $123$ vorton for various values of $\kappa$ and $\beta$.}
        \label{fig:mwvort123}
\end{figure}
%As previously noted, the lensing effect like the Einstein ring in \ref{fig:lensimcvortk0b05mw}, just as the readily apparent symmetry in \ref{fig:gridcircns123}. The discontinuity on the string (projected onto the image plane) is also apparent. The other generic properties of lensing image like shear is also present in all lensing images from all type of vorton and cosmic string discussed here. However, we note that in order to observe these discontinuity, the vorton ought to have at least a galaxy behind it, and that the vorton is of the angular size resolvable to telescope.

%\section{Conclusions}

%In our previous work~\cite{Putra:2024zms}, we analyzed the lensing patterns of circular chiral vortons by deriving their metric and studying the associated null geodesics. In this study, we have extended the analysis to non-circular configurations. Specifically, we derived stationary solutions to the Nambu–Goto equations describing vorton loops with arbitrary harmonic modes and investigated their gravitational lensing effects. While this generalization sacrifices an explicit metric solution, we adopted the thin-lens approximation, treating the vorton’s contribution to the spacetime geometry as a weak perturbation on a flat background~\cite{deLaix:1996vc}.

Our findings reveal several distinctive features of vorton lensing. In the circular case, the images exhibit an apparent discontinuity, separating minimally distorted regions from highly distorted ones, both originating from the same source. Remarkably, an Einstein ring can form while the original source image remains simultaneously visible at the center of the ring with only mild distortion (see Fig.\ref{fig:mwcvortn200}). This unusual coexistence of an Einstein ring and a nearly undistorted central source may be a phenomenon unique to cosmic string loops, offering a potential observational signature. For non-circular vortons, however, the central image becomes more distorted, as shown in Fig.\ref{fig:mwvort123}, indicating a stronger interplay between shape deformations and lensing patterns.

Another notable feature arises from the effects of frame dragging. For certain configurations (Figs.~\ref{fig:gridcircnscvort}, \ref{fig:gridcircnsturok}, and \ref{fig:gridcircns123}), the frame-dragging effect breaks the symmetry of lensing images, in stark contrast with ordinary Nambu–Goto strings, which generate symmetric double-image patterns due to the absence of rotation. This asymmetry makes vortons distinguishable through direct lensing observations, and for vortons with sufficiently large angular size, gravitational lensing may serve as a promising detection channel.

From a broader perspective, it is useful to put vorton lensing alongside more common astrophysical lensing phenomena. Black holes and compact objects typically produce highly symmetric Einstein rings or multiple images with characteristic magnification patterns~\cite{Virbhadra:1999nm, Bozza:2010xqn}. Galaxy- or cluster-scale lenses give rise to arcs, arclets, or strong shear fields~\cite{Treu:2010uj}. Ordinary cosmic strings generate a pair of identical, undistorted images separated by an angular deficit~\cite{Vilenkin:1984ea}. Put in this context, vorton lensing is distinctive: it can simultaneously produce (i) an Einstein ring plus a nearly undistorted central source, (ii) discontinuous transitions in image distortion across angular sectors, and (iii) asymmetric image patterns induced by frame dragging. Some degeneracies are nevertheless possible: for instance, vorton-induced asymmetric rings may be confused with lensing by rotating compact objects or binary lenses~\cite{Bozza:2002zj}, while discontinuous distortions could mimic effects of substructure in galaxy lenses. However, the coexistence of a sharp discontinuity, central image survival, and frame-dragging asymmetry provides a clearer path to distinguishing vortons from these more conventional lenses.

Despite their distinct signatures, it is important to consider cosmological implications of vortons. A longstanding concern is the vorton abundance problem. Early studies indicated that stable vortons, once produced in the early Universe, could survive indefinitely and potentially overclose the Universe or exceed current bounds on the dark matter density~\cite{Brandenberger:1996zp}. Several mechanisms have been proposed to mitigate this issue, including current leakage, electromagnetic radiation, or plasma interactions that destabilize loops before they reach cosmological abundances~\cite{Peter:1992dw, Martins:1998gb, Carter:1999an}. More realistic field-theoretic treatments, such as including fermion backreaction or higher-order corrections, may further reduce their long-term stability~\cite{Battye:2008mm}.

On the other hand, if vortons are sufficiently long-lived but produced at a subdominant rate, they could constitute a fraction of the dark matter in the Universe~\cite{Davis:1988jq, Brandenberger:1996zp, Carter:1999an}. In this scenario, vortons would behave as macroscopic non-relativistic relics, with potentially observable consequences in gravitational microlensing searches, cosmic microwave background constraints, or stochastic gravitational wave backgrounds~\cite{Auclair:2019wcv, Battye:2008mm}. Thus, while their abundance must be carefully constrained, the possibility that vortons might serve as exotic dark matter candidates remains an intriguing topic for research.

Finally, we note that our construction of non-circular vortons has thus far been limited to the chiral-current case. Whether other superconducting string models, such as Witten’s original bosonic current model~\cite{Witten:1984eb}, can also yield stable non-circular vorton solutions remains an open question. Addressing this, along with a more detailed cosmological treatment of vorton formation and abundance, would be natural extensions of the present work.

\label{sec:conc}

\acknowledgments

We thank Steven Holme and Jose Blanco-Pillado for the discussions on the dynamics of vortons and its cosmological signatures, respectively. This work is funded by the Hibah PUTI Q1 UI No.~PKS-196/UN2.RST/HKP.05.00/2025.

\bibliographystyle{agsm}

\end{document}